\documentstyle[]{mn}
\newif\ifAMStwofonts
\ifoldfss
  \newcommand{\rmn}[1] {{\rm #1}}

  \ifCUPmtlplainloaded \else
    \NewTextAlphabet{textbfit} {cmbxti10} {}
    \NewTextAlphabet{textbfss} {cmssbx10} {}
    \NewMathAlphabet{mathbfit} {cmbxti10} {} % for math mode
    \NewMathAlphabet{mathbfss} {cmssbx10} {} %  "   "    "
  \fi
  \ifAMStwofonts
    \ifCUPmtlplainloaded \else
      \NewSymbolFont{upmath} {eurm10}
      \NewSymbolFont{AMSa} {msam10}
      \NewMathSymbol{\upi}     {0}{upmath}{19}
      \NewMathSymbol{\umu}     {0}{upmath}{16}
      \NewMathSymbol{\upartial}{0}{upmath}{40}
      \NewMathSymbol{\leqslant}{3}{AMSa}{36}
      \NewMathSymbol{\geqslant}{3}{AMSa}{3E}

      \let\leq=\leqslant 
      \let\geq=\geqslant 
    \fi
  \fi
\fi % End of OFSS

\ifnfssone
  \newmathalphabet{\mathit}
  \addtoversion{normal}{\mathit}{cmr}{m}{it}
  \addtoversion{bold}{\mathit}{cmr}{bx}{it}
  \newcommand{\rmn}[1] {\mathrm{#1}}

  \newmathalphabet{\mathbfit} % math mode version of \textbfit{..}
  \addtoversion{normal}{\mathbfit}{cmr}{bx}{it}
  \addtoversion{bold}{\mathbfit}{cmr}{bx}{it}
  \newmathalphabet{\mathbfss} % math mode version of \textbfss{..}
  \addtoversion{normal}{\mathbfss}{cmss}{bx}{n}
  \addtoversion{bold}{\mathbfss}{cmss}{bx}{n}
  \ifAMStwofonts
    \ifCUPmtlplainloaded \else
      \UseAMStwoboldmath
      \makeatletter
      \new@mathgroup\upmath@group
      \define@mathgroup\mv@normal\upmath@group{eur}{m}{n}
      \define@mathgroup\mv@bold\upmath@group{eur}{b}{n}
      \edef\UPM{\hexnumber\upmath@group}
      \new@mathgroup\amsa@group
      \define@mathgroup\mv@normal\amsa@group{msa}{m}{n}
      \define@mathgroup\mv@bold\amsa@group{msa}{m}{n}
      \edef\AMSa{\hexnumber\amsa@group}
      \makeatother
      \mathchardef\upi="0\UPM19
      \mathchardef\umu="0\UPM16
      \mathchardef\upartial="0\UPM40
      \mathchardef\leqslant="3\AMSa36
      \mathchardef\geqslant="3\AMSa3E

      \let\leq=\leqslant 
      \let\geq=\geqslant 
    \fi
  \fi
\fi % End of NFSS release 1

\ifnfsstwo
  \newcommand{\rmn}[1] {\mathrm{#1}}

  \DeclareMathAlphabet{\mathbfit}{OT1}{cmr}{bx}{it}
  \SetMathAlphabet\mathbfit{bold}{OT1}{cmr}{bx}{it}
  \DeclareMathAlphabet{\mathbfss}{OT1}{cmss}{bx}{n}
  \SetMathAlphabet\mathbfss{bold}{OT1}{cmss}{bx}{n}
  \ifAMStwofonts
    \ifCUPmtlplainloaded \else
      \DeclareSymbolFont{UPM}{U}{eur}{m}{n}
      \SetSymbolFont{UPM}{bold}{U}{eur}{b}{n}
      \DeclareSymbolFont{AMSa}{U}{msa}{m}{n}
      \DeclareMathSymbol{\upi}{0}{UPM}{"19}
      \DeclareMathSymbol{\umu}{0}{UPM}{"16}
      \DeclareMathSymbol{\upartial}{0}{UPM}{"40}
      \DeclareMathSymbol{\leqslant}{3}{AMSa}{"36}
      \DeclareMathSymbol{\geqslant}{3}{AMSa}{"3E}

      \let\leq=\leqslant 
      \let\geq=\geqslant 
    \fi
  \fi
\fi % End of NFSS release 2

\ifCUPmtlplainloaded \else
  \ifAMStwofonts \else % If no AMS fonts
    \def\upi{\pi}
    \def\umu{\mu}
    \def\upartial{\partial}
  \fi
\fi

\newcommand{\unit}[1]
        {{\mbox{\rm\,\,#1}}}

\newcommand{\magarcsec}
        {\unit{mag arcsec$^{-2}$}}

\newcommand{\MByte}
        {MB}
\newcommand{\GByte}
        {GB}

\newcommand{\grs}
        {GRS}
\newcommand{\grss}
        {GRSs}

\newcommand{\AAO}
        {AAO}

\newcommand{\TdF}
        {2dF}

\newcommand{\AAT}
        {AAT}

\newcommand{\UKST}
        {UKST}
\newcommand{\SDSS}
        {SDSS}
\newcommand{\NASA}
        {NASA}

\newcommand{\APM}
        {APM}
\newcommand{\psf}
        {PSF}

\newcommand{\psfs}
        {PSFs}

\newcommand{\samethanks}
	{{\Huge $^\star$}}

\newcommand{\vect}[1]
        {\mbox{\boldmath ${#1}$}}

\newcommand{\etc}
	{etc.}
\newcommand{\etal}
	{et al.}
\newcommand{\eg}
	{e.g.}
\newcommand{\cf}
	{c.f.}
\newcommand{\ie}
	{i.e.}
\newcommand{\eq}[1]
	{equation~(\ref{equation:#1})}

\newcommand{\sect}[1]
	{Section~\ref{section:#1}}

\newcommand{\sects}[1]
        {Sections~\ref{section:#1}}

\newcommand{\fig}[1]
	{Fig.~\ref{figure:#1}}
\newcommand{\figs}[1]
        {Figs.~\ref{figure:#1}}
\newcommand{\Fig}[1]
        {Fig.~\ref{figure:#1}}

\newlength{\singlefigureheight}
\newlength{\doublefigureheight}
\newlength{\triplefigureheight}
\newlength{\squarefigureheight}
\newcommand{\AaA}
        {A\&A}

\newcommand{\AJ}
        {AJ}
\newcommand{\ApJ}
        {ApJ}
\newcommand{\ApJS}
        {ApJS}

\newcommand{\ARAA}
        {ARA\&A}

\newcommand{\MNRAS}
        {MNRAS}
\newcommand{\PhilTransA}
        {Phil.\ Trans.\ of the Royal Soc.\ A}

\newcommand{\PASA}
        {PASA}

\newcommand{\FL}
        {FL}

\begin{document}

\title[Lensing in the 2dF redshift survey]
{Using the 2dF galaxy redshift survey to 
detect gravitationally-lensed quasars}

\author[D.\ J.\ Mortlock and R.\ L.\ Webster]
       {
        Daniel J.\ Mortlock$^{1,2,3}$\thanks{
		E-mail:
        	mortlock@ast.cam.ac.uk (DJM); 
		rwebster@physics. ph.unimelb.edu.au (RLW)}
	and Rachel L.\ Webster$^1$\samethanks\ \\
        $^1$School of Physics, The University of Melbourne, Parkville,
        Victoria 3052, Australia \\
        $^2$Astrophysics Group, Cavendish Laboratory, Madingley Road,
        Cambridge CB3 0HE, U.K. \\
        $^3$Institute of Astronomy, Madingley Road, Cambridge
        CB3 0HA, U.K. \\
       }

\date{
Accepted 
Received;
in original form 2000 June 5}

\pagerange{\pageref{firstpage}--\pageref{lastpage}}
\pubyear{2000}

\label{firstpage}

\maketitle

\begin{abstract}
Galaxy redshift surveys can be 
used to detect gravitationally-lensed quasars if the 
spectra obtained are searched for the quasars' emission lines. 
Previous investigations of this possibility have used 
simple models to show that the 2 degree Field (\TdF) redshift survey
could yield several tens of new lenses, and that the 
larger Sloan Digital Sky Survey should contain an order of magnitude more.
However the particular selection effects of the samples were not
included in these calculations, limiting the robustness of the 
predictions; 
thus a more detailed simulation of the \TdF\ survey 
was undertaken here.
The use of an isophotal magnitude limit reduces both the
depth of the sample and the expected number of lenses,
but more important is the Automatic Plate Measuring survey's
star-galaxy separation algorithm, used to generate the \TdF\
input catalogue.
It is found that most quasar lenses are classed as merged stars,
with only the few lenses with low-redshift deflectors
likely to be classified as galaxies.
Explicit inclusion of these selection effects 
implies that the \TdF\ survey should contain 
10 lenses on average.
The largest remaining uncertainty is the lack of knowledge of the ease
with which any underlying quasars can be extracted from the survey spectra.
\end{abstract}


\section{Introduction}
\label{section:intro_2df}

Gravitationally-lensed quasars are extremely valuable, 
and so a great deal of effort is expended searching 
for new lens systems. 
Most lens surveys are based on high-resolution 
re-observation of known quasars, but lensed quasars
can also be found galaxy redshift surveys (\grss)
if the spectra obtained are searched for the presence
of quasar emission lines.
As well as being efficient, this form of lens search 
is complementary to the conventional 
surveys, being primarily sensitive to low-redshift deflectors --
the galaxies that would enter the \grs\ a priori.
This is important as the proximity of the lens 
galaxy can facillitate a number of unique observations,
as exemplified by Q~2237+0305 (Huchra \etal\ 1985),
the only quasar lens yet found in a \grs.

A more detailed discussion of these points is 
contained in Mortlock \& Webster (2000b), 
one of the two previous investigations of quasar
lensing in redshift surveys. 
An `initial, order of magnitude estimate' of the 
number of lenses expected in future \grss\ by Kochanek (1992) 
suggested there would be one lens per $\sim 10^5$ galaxy spectra,
but this was shown to be an under-estimate by up to an order of 
magnitude, as the combined light of the quasar images and the 
lens galaxy must be included in the calculation (Mortlock \& Webster 2000b).
Both these these results were obtained 
for an idealised \grs, characterised only by the number of 
spectra obtained and a survey's magnitude\footnote{All magnitudes are
in the $B_{\rmn J}$ system, but this subscript is omitted
for brevity. Also, $m$ denotes total apparent magnitude,
$m^\prime$ isophotal apparent magnitude (as defined in \sect{apm survey}),
and $M$ denotes absolute magnitude.
An Einstein-de Sitter cosmologcal model is assumed for simplicity;
the choice cosmology does not have a significant impact
on the inferred properties of the \TdF\ galaxy survey 
or the lens statistics, as the galaxies are so nearby (Kochanek 1992).
Finally, Hubble's constant is taken to be 70 km s$^{-1}$ Mpc$^{-1}$.} limit.
It is important to test the robustness of these simple predictions,
which can be done most profitably 
by simulating a real survey in detail.
The two obvious candidates are 
the 2 degree Field (\TdF) galaxy survey 
(\eg\ Colless 1999; Folkes \etal\ 1999)
and 
the Sloan Digital Sky Survey 
(\SDSS; \eg\ Szalay 1998; Loveday \& Pier 1998).
The former was chosen primarily as it subject to a number of 
comlex, surface brightness-related selection effects, whereas the 
\SDSS\ is much closer to the ideal galaxy sample modelled in 
Mortlock \& Webster (2000b). Furthermore, the \TdF\ survey will be completed
first, offering the most immediate opportunity to discover new 
quasar lenses in a galaxy survey.

The selection criteria of the \TdF\ \grs\ and its 
parent survey are carefully defined in 
\sect{2df}. 
The main aim of these selection criteria is to 
exclude stars (\sect{stars_2df})
whilst maximising the completeness of the galaxy
sample (\sect{galaxy_2df}), although care was also taken 
to reject unwanted merged images (\sect{merge}).
In order to determine the number of lenses in the survey, 
the same selection effects must be applied to a simulated lens
population,
and the spectral sensitivity of the lens search must also
be included (\sect{lens_2df}).
The results of the this calculation 
and the relative importance of 
the various survey characteristics
are given in \sect{results_2df}.
The main conclusions are summarised in \sect{conc_2df}.

\section{The 2dF galaxy redshift survey}
\label{section:2df}

The \TdF\ instrument (\sect{the 2df}) is used for 
a number of projects. The largest is 
the eponymous galaxy survey (\sect{survey}), the
properties of which are, however, determined mainly 
by its parent sample (\sect{apm survey}).

\subsection{The 2dF instrument}
\label{section:the 2df}

The \TdF\ instrument (Taylor 1994; Cannon 1995)
is a wide-field, multi-fibre spectroscopic 
survey facility for the Anglo-Australian Telescope (\AAT).
Corrective optics give aberration-free images over
the full 2~deg-diameter field of view, but the main innovation
is the automatically-configured 
400-fibre spectrograph in the prime focus plane of the 
telescope. The positional accuracy of 
$0.1$ arcsec is required because
the fibres have an effective angular radius of only
$\theta_{\rmn f} = 1$ arcsec.
This aperture size is comparable to the seeing at the site --
which varies between
1 arcsec and 5 arcsec, with a median value of $\sim 2$ arcsec
(Ryan \& Wood 1995) -- and so the spectra obtained are 
inevitably of variable quality.

\subsection{The \TdF\ survey}
\label{section:survey}

The \TdF\ \grs\ (\eg\ Colless 1999; Folkes \etal\ 1999)
is the largest single project
that is proposed for the \TdF\
instrument\footnote{The \TdF\ quasar survey (\eg\ Boyle \etal\ 1999a,b)
is a separate project, but data for both it and the \grs\
are obtained simultaneously.}.
The survey
will require $\sim 100$ nights of observing time
to obtain the redshifts of $N_{\rmn tot} = 2.5 \times 10^5$ objects,
about 95 per cent of which are expected to be galaxies.
The spectrograph being used has a resolution of 
9 \AA\ from $\sim 3700$ \AA\ to $\sim 8100$ \AA, 
and 45-minute integrations
should yield spectra with a signal-to-noise ratio of $\sim 20$ per
pixel at the survey magnitude
limit of $m^\prime \simeq 19.5$. 

In the absence of any other selection effects,
the survey limit, together with $N_{\rmn tot}$,
%and the efficiency, $E \geq 0.95$, 
would completely characterise the survey,
and it could then be simulated using standard techniques.
However, the catalogue of galaxy candidates is taken from the 
Automatic Plate Measuring (\APM)
galaxy survey, which is subject to a number of 
complex, surface brightness-related
selection effects. 

\subsection{The \APM\ galaxy survey}
\label{section:apm survey}

The \APM\ system (\eg\ Kibblewhite \etal\ 1984), 
%located in Cambridge, \UK, 
combines a high-speed microdensitometer 
with a dedicated computer system capable of on-line
image detection. 
Used to analyse United Kingdom Schmidt Telescope (\UKST) photographic plates,
it can provide spatial information on a pixel scale of $\sim 0.5$ arcsec,
and relative photometry over several orders of magnitude in flux. 
Scanned at this resolution,
a typical \UKST\ plate
represents $\sim 3$ \GByte\ of information, but the image detection
algorithms allow it to be stored as a $\sim 3$ \MByte\ list of image parameters
with minimal information loss.
The \APM\ software defines an image as any connected set of 16 or more 
pixels which are all above the surface brightness limit.
This limit varies between $\mu_{\rmn lim} = 24$ mag arcsec$^{-2}$
and $\mu_{\rmn lim} = 25$ mag arcsec$^{-2}$. 
The minimum image area of 
$\sim 4$ arcsec$^2$ then implies an isophotal
magnitude limit of $m^\prime \sim 22$.
For each image, 15 parameters (defined in Maddox \etal\ 1990a) are
calculated from the sky-subtracted plate densities; these comprise the
standard output of the \APM\ system.

The \APM\ galaxy survey
(Maddox \etal\ 1990a; 
Maddox, Efsathiou \& Sutherland 1990b;
Dalton \etal\ 1994;
Maddox, Efsathiou \& Sutherland 1996; 
Dalton \etal\ 1997)
is a uniform sample of $\sim 2 \times 10^6$ galaxies 
to a magnitude limit of $m^\prime = 20.5$.
It was compiled from \APM\ scans of 185 contiguous \UKST\ survey plates,
and the galaxy sample was thus selected 
from $\sim 2 \times 10^7$ images brighter than 
$m^\prime \simeq 22$ spread over a total area of 4300 deg$^2$. 

\subsubsection{Star-galaxy separation}

Given that only a third of objects brighter
than $m \simeq 20.5$ are galaxies, one of the most 
important aspects of the \APM\ galaxy survey is the 
star-galaxy separation technique.
In Mortlock \& Webster (2000b), image ellipticity was used as a generic
method of discriminating between point-like images and 
extended sources --
this was useful as a simple estimate, but more sophisticated
methods are required to construct a galaxy catalogue from 
real two-dimensional data. Most star-galaxy separation techniques
(\eg\ Jarvis \& Tyson 1981; 
Godwin, Metcalfe \& Peach 1983; 
Maddox \etal\ 1990a)
are based on two facts: 
a galaxy has a lower peak surface brightness 
than a star of the same magnitude;
and the surface brightness profile of a star is determined 
purely by observational effects.
Stars and galaxies inhabit different regions of the phase-space
defined by magnitude, area, peak surface brightness, \etc, 
and scatter plots of the data can be used to separate the two 
populations.

The \APM\ galaxy sample was defined in a more
complex manner, using the areal profiles of each object as the 
main diagnostic.
Amongst the image parameters provided by the \APM\ is
the area of each image above eight plate density levels, $A_i$,
where $1 \leq i \leq 8$. 
The plate density is proportional to (logarithmic) surface brightness
fainter than $\sim 20$ mag arcsec$^{-2}$
(Maddox \etal\ 1990a), 
so the levels can be approximated by
\begin{equation}
\mu_i = \mu_{\rmn lim} + i \, \Delta \mu,
\end{equation}
where $\mu_{\rmn lim} \simeq 24.5 \magarcsec$ 
and $\Delta \mu \simeq 0.75$ mag arcsec$^{-2}$.
This approximation slightly underestimates $\mu_8$, due to 
saturation effects, and also ignores plate-to-plate
variations.
The random errors in the $A_i$s were found to be well 
approximated by the Poisson error in the number of pixels 
above the relevant surface brightness. 
Given the \APM\ scale of $0.5$ arcsec per pixel, 
$\Delta A_i \simeq 2 \sqrt{A_i}$ if $A_i$ is expressed
in arcsec$^2$.
The areal profile of an image can be thought of as a plot of
area versus surface brightness, which is similar 
to a conventional surface brightness profile.
As shown in Maddox \etal\ (1990a), stars have a characteristic 
areal profile at any given magnitude, which is distinct from 
the more varied profiles of galaxies. However this differentiation
breaks down for $m^\prime \ga 20$ (as there is 
limited information for such faint images)
and for $m^\prime \la 17$ (as the stars appear extended due to
saturation effects).

As a source of extra information, the peak surface brightness
of each image, $\mu_{\rmn peak}$, and the radius of 
gyration,
$\theta_{\rmn G}$, were also used, giving a total of 10 parameters for
the star-galaxy separation.
The radius of gyration is defined as 
\begin{equation}
\label{equation:th_gyr}
\theta_{\rmn G} = \sqrt{\langle \theta_x^2 \rangle
+ \langle \theta_y^2 \rangle},
\end{equation}
where the expectation values are given by surface
brightness-weighted integrals over the 
area of the image for which $\mu \leq \mu_{\rmn lim}$.
The uncertainties in $\mu_{\rmn peak}$ and $\theta_{\rmn G}$ are not given in
Maddox \etal\ (1990a), but are estimated to be $\sim 10$ per cent.
As with the $A_i$s, a distinct stellar locus is apparent
for both $\mu_{\rmn peak}$ and $\theta_{\rmn G}$ as a function of 
magnitude -- the stars have higher peak surface brightnesses and 
smaller radii of gyration than the galaxies.

The final image classification statistic
used by Maddox \etal\ (1990a) combines all 10 image parameters
in a $\chi^2$-like formulation.
For an object of isophotal magnitude $m^\prime$, it is defined as
\begin{eqnarray}
\label{equation:psi_apm}
\psi_{\rmn APM} & \!\! = \!\! & {\rmn 2000} \log \left\{ \frac{1}{10}\left\{
\sum_{i = 1}^8 \left[\frac{A_i - A_{i,{\rmn s}} (m^\prime)}
{\sqrt{4 A_{i,{\rmn s}} (m^\prime)}}\right]^2 \right. \right. \\
& \!\! + \!\! & \left. \left. \!\!
\left[\frac{\theta_{\rmn G} - \theta_{\rmn G,{\rmn s}} (m^\prime)}
{\sqrt{2 \theta_{\rmn G,{\rmn s}} (m^\prime)}}
\right]^2 
+ \left[ \frac{f_{\rmn peak} - f_{\rmn peak,s} (m^\prime)}
{\sqrt{f_{\rmn peak,s} (m^\prime)}} \right]^2\right\} \!\!
\right\}, \nonumber
\end{eqnarray}
where 
the $A_i$ are in units of arcsec$^2$,
$\theta_{\rmn G}$ is in units of arcsec,
and $f_{\rmn peak}$ is the peak surface brightness
in linear units, defined to match the \APM\ plate
density.
The `s' subscript denotes the parameter value of 
an isolated star of the same isophotal magnitude
as the image in question.
Each term is normalised by the expected 
uncertainty in the stellar locus, 
as in a $\chi^2$ formulation.
Note that the definition of $\psi_{\rmn APM}$ 
given in Maddox \etal\ (1990a) does not include 
the overall normalisation (\ie\ the `1/10' term); 
it is included here to ensure that stellar locus 
is centred on $\psi_{\rmn APM} = 0$. The
prefactor is changed from 1000 to 2000 to
compensate for this.

\begin{figure}
\includegraphics{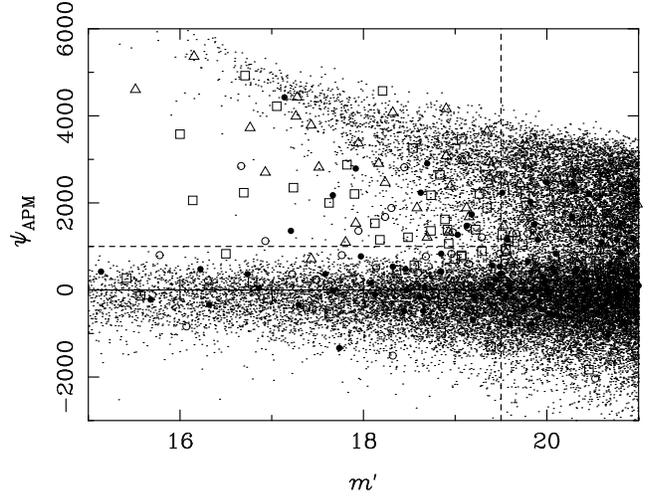}
\vspace{\singlefigureheight}
\caption{
The primary \APM\ star-galaxy separation parameter, $\psi_{\rm APM}$,
as a function of \APM\ isophotal magnitude, $m^\prime$.
The stars (points) lie along $\psi_{\rm APM} = 0$, by definition, and
the galaxies (also points) comprise the well-defined `ridge'
with $\psi_{\rm APM} > $
1000 (above the horizontal dashed 
line), although the distinction between
the two populations is less clear at fainter magnitudes.
Also shown are representative populations of lensed quasars (solid
circles) and pairs of merged images:
star-star mergers (open circles); star-galaxy mergers (open
squares) and galaxy-galaxy mergers (open triangles). The vertical
dashed line at $m^\prime = 19.5$ represents the \TdF\ survey limit.}
\label{figure:psi_apm}
\end{figure}

\Fig{psi_apm} shows a plot of $\psi_{\rmn APM}$ as a function of
isophotal magnitude for stars, galaxies and mergers, and
is directly comparable to Fig.\ 10 of Maddox \etal\ (1990a).
Some of the details differ (\eg\ the exact position and width of 
the stellar locus and absence of saturation effects in the 
simulation), 
but the generic features are the same. 
The stars (See \sect{stars_2df}.) are centred on $\psi_{\rmn APM} = 0$, 
and their spread is independent of the point-spread function (\psf)
due to the normalisations in \eq{psi_apm}.
The galaxies (See \sect{galaxy_2df}.) lie along a `ridge' 
with $\psi_{\rmn APM} \ga$ 1000, but are less distinct from 
the stars at fainter magnitudes.
However $\psi_{\rmn APM}$ cannot be used to separate merged 
images from extended objects; further specialised statistics 
are required.

\subsubsection{Merged objects}

Images separated by 
$\la 5$ arcsec are registered as single objects by
the \APM\ software, and tend to have 
$\psi_{\rmn APM} \ga$ 1000, as illustrated in \fig{psi_apm}.
If this was the only star-galaxy separation implemented,
most lensed quasars would be included in 
the \APM\ galaxy survey.
However Maddox \etal\ (1990a) 
used the radius of gyration and a saturation parameter to 
detect (and, in general, reject) merged images.

The radius of gyration, $\theta_{\rmn G}$, as defined 
in \eq{th_gyr} is a weighted measure of 
the extent of an image, and is evaluated on a 
pixel-by-pixel basis by the \APM\ software. 
If an image is assumed to be perfectly elliptical, 
an approximation to the radius of gyration is given by
\begin{equation}
\theta^\prime_{\rmn G} = 
\left[
\frac{\sum_{i = 1}^7 f_i A_i (A_i - A_{i + 1})}
{2 \pi \sqrt{1 - e^2} \sum_{i = 1}^7 f_i (A_i - A_{i + 1})}
\right]^{1/2},
\end{equation}
where $e$ is the overall measured ellipticity of the image
and $f_i$ the surface brightness of the $i$th level
in linear units
(Maddox \etal\ 1990a).
For isolated images, 
$\theta_{\rmn G} \simeq \theta_{\rmn G}^\prime$, with some 
scatter due to random noise; 
for multiple images $\theta_{\rmn G} > \theta_{\rmn G}^\prime$,
as dumbbell-shaped image pairs have higher second
order moments than isolated images with the same areal profile.
From this Maddox \etal\ (1990a) define 
\begin{equation}
k_{\rmn APM} = \frac{\theta_{\rmn G}^2 / \theta_{\rmn G}^{\prime 2}}
{\theta_{\rmn G,s}^2(m^\prime) / \theta_{\rmn G,s}^{\prime 2}(m^\prime)}
\simeq \frac{\theta_{\rmn G}^2}{\theta_{\rmn G}^{\prime 2}},
\end{equation}
where again the `s' subscript denotes stellar values at
the same isophotal magnitude as the (merged) object in question.
If $k_{\rmn APM}$ is much greater than unity the object
is quite likely to be a merged pair of images or
a higher multiple.

The $k$-parameter is sufficient to find pairs of nearby images,
but is not sensitive to superimposed objects (\eg\ a 
star centred on a galaxy) and cannot be used to differentiate between
galaxy-galaxy mergers, star-galaxy mergers and 
star-star mergers. 
Such a distinction can be made by looking at the
fraction of the image that is close to saturation on 
the \UKST\ plates.
To this end Maddox \etal\ (1990a) defined 
\begin{equation}
\label{equation:mu_apm}
\mu_{\rmn APM} = 
\frac{\left(0.2 A_7 + A_8 \right) / A_1}
{\left[0.2 A_{7, {\rmn s}} (m^\prime) + A_{8, {\rmn s}} (m^\prime) \right]
/ A_{1,{\rmn s}} (m^\prime)},
\end{equation}
where the inclusion of $A_7$ (rather than just using $A_8/A_1$)
gives extra stability for fainter images.
By definition stars (both single and multiple)
have $\mu_{\rmn APM} \simeq 1$,
whereas galaxies (and galactic mergers)
$0 \leq \mu_{\rmn APM} < 1$.

\begin{figure}
\includegraphics{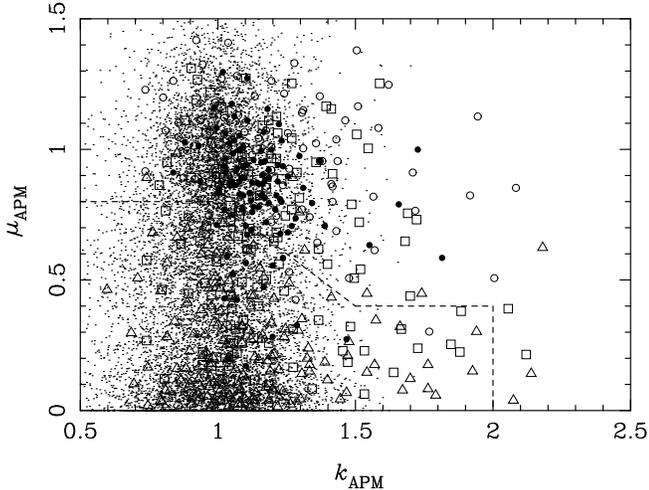}
\vspace{\singlefigureheight}
\caption{A scatter plot of the two parameters
used to characterise 
merged images in the \APM\ galaxy survey, $k_{\rm APM}$ and $\mu_{\rm APM}$.
The dashed line defines the boundary between `stellar' mergers 
(above the line) and `galactic' mergers (below the line) --
only the latter are included in the galaxy survey.
The stars (points) are centred at $k_{\rm APM} = 1$
and $\mu_{\rm APM} = 1$, by definition.
The galaxies (also points) have $k_{\rm APM} \simeq 1$ as well,
but a broad spread in $\mu_{\rm APM}$.
Also shown are representative populations of lensed quasars (solid
circles) and pairs of merged images:
star-star mergers (open circles); star-galaxy mergers (open
squares) and galaxy-galaxy mergers (open triangles).}
\label{figure:k-mu}
\end{figure}

A scatter plot in $k_{\rmn APM}$-$\mu_{\rmn APM}$ 
space (\fig{k-mu}; \cf\ Fig.~14 of Maddox \etal\ 1990a) 
can thus be used to separate 
the various types of mergers from isolated stars and galaxies.
As explained in Maddox \etal\ (1990a), merged galaxies
were retained, as were bright galaxies blended 
with fainter stars, but all other stellar mergers were rejected.
This was done by making a cut `by eye' in the 
$k_{\rm APM}$-$\mu_{\rm APM}$ plane, which is shown as the 
dashed line in \fig{k-mu}.
Note also that the normalisation of both 
$k_{\rmn APM}$ and $\mu_{\rmn APM}$ is such that there
is no intrinsic magnitude dependence in the merger
classification.

\section{Stars}
\label{section:stars_2df}

As the \APM\ image classification scheme is 
defined relative to the stellar population, stars
represent a control sample. 
Misclassified stars are also the primary source of 
contamination in the galaxy catalogue, 
as they represent such a large fraction of sources 
over the magnitude range of the survey.

The model of the stellar population used was described in 
Mortlock \& Webster (2000b). Briefly, 
the number counts are taken from Bahcall \& Soneira (1980)
and the images are assumed to have 
a Moffat (1969) \psf, defined by
\begin{equation}
\label{equation:moffat}
f_{\rmn see}({\theta}) =
\frac{(\eta - 1) (2^{1/\eta} - 1)}
{\pi (\theta_{\rmn{s}} / 2)^2}
\left[1 + (2^{1/\eta} - 1)
\frac{\theta^2}
{(\theta_{\rmn{s}} / 2)^2}\right]^{-\eta},
\end{equation}
where $\theta_{\rmn s}$ is the full width at half-maximum
of the seeing disc and $\theta$ is angular position on
the sky.
As discussed in Mortlock \& Webster (2000b), two 
values of $\eta$ are used: $\eta = 2.5$, the default
value assumed by the
{\sc image reduction and analysis facility}
software
(Tody 1986); and 
$\eta \rightarrow \infty$, a Gaussian \psf.
It is important to note that this \psf\
is formally defined for an object of a given
total, as opposed to isophotal magnitude.
An iterative procedure is required to find the total
magnitude, $m$, which corresponds to a given isophotal 
magnitude $m^\prime$.
The random errors due to the finite signal-to-noise 
of the \APM\ pixels are not explicitly simulated
here; only the approximately Gaussian errors discussed 
in \sect{apm survey} are included.

\Fig{psi_apm} shows the stellar locus as the `ridge' 
with $\psi_{\rmn APM} \simeq 0$.
The spread is independent of the \psf\ as $\psi_{\rm APM}$
is normalised relative to the spread of the stellar 
images.
In reality the stellar locus is not straight
and has non-Gaussian outliers (\cf\ Fig.\ 10 of 
Maddox \etal\ 1990a), 
but the most important features are present.

Isolated stars have an even simpler distribution on 
the $k_{\rmn APM}$-$\mu_{\rmn APM}$ plot (\fig{k-mu})
used to 
distinguish between mergers and isolated objects.
Their distribution in $k_{\rmn APM}$ is simply 
a Gaussian with spread determined by the 
assumed error in the radius of gyration discussed
in \sect{apm survey}. 
The use of $A_7$ and $A_8$ in the definition of 
$\mu_{\rmn APM}$ [\eq{mu_apm}] means that the 
$\mu_{\rmn APM}$ distribution is marginally magnitude dependent, 
but is also close to a Gaussian for stars 
with $\mu_{\rmn peak} \simeq \mu_8$.

\section{Galaxies}
\label{section:galaxy_2df}

The main purpose of the morphological classification criteria
of the \APM\ survey is to generate a reasonably complete 
and contamination-free sample of galaxies. 
The properties of the galaxy sample are determined by
the intrinsic galaxy population (\sect{pops_2df})
and their observed surface brightness profiles 
(\sect{sb_obs}). The latter dominates both the \APM\ 
classifications (\sect{class_2df}) and 
the \TdF\ fibre magnitudes (\sect{m_fib_g}), but the two 
combine to determine the redshift distribution of
the survey galaxies (\sect{dpdz_2df}).

\subsection{Populations}
\label{section:pops_2df}

The model of the local galaxy population is described in 
detail in Mortlock \& Webster (2000b), and thus 
only outlined here.
Three galaxy types are considered: spirals (S) 
and two classes of elliptical galaxies (E and S0), 
the relative numbers of which are given in Postman \& Geller (1984).
Each population is assumed to follow the Schechter (1976) luminosity 
functions detailed in Efstathiou, Ellis \& Peterson (1988).
(Thus $M_* = - 19.4 \pm 0.1$ for all three types.)
In the calculation of the galaxies' lensing properties 
(\sects{lens_2df} and \ref{section:results_2df})
they are modelled as isothermal spheres 
(\eg\ Turner, Ostriker \& Gott 1984; Binney \& Tremaine 1987),
with the standard Faber-Jackson (1976) and
Tully-Fisher (1977) relations used to
convert luminosities to velocity dispersions.

\subsection{Surface brightness profiles}
\label{section:sb_obs}

In the \APM\ survey, galaxies are distinguished from stars 
primarily by their surface brightness profiles. 
In particular, the ellipticity or elongation of images is
not used in the object classification\footnote{Ellipticity 
was the only criterion used for star-galaxy separation in 
Mortlock \& Webster (2000b).}, and so
all galaxies can be assumed to have circular isophotes 
without loss of generality.

In the absence of atmospheric seeing, 
the surface brightness of galaxies is given by their
intrinsic (radial) surface brightness profiles.
Spirals are assumed to follow a Freeman (1970)
profile, given by
\begin{equation}
f_{\rmn S} (\theta) = \frac{F}{2 \pi (0.596\, \theta_{\rmn g})^2} \,
e^{- \theta / (0.596 \, \theta_{\rmn g})} ;
\end{equation}
the two types of ellipticals are modelled by
a de Vaucouleurs (1948) profile,
given by
\begin{equation}
f_{\rmn E} (\theta) = 296.7 \frac{F}{\pi \theta_{\rmn g}^2} \,
e^{- 7.67 (\theta / \theta_{\rmn g})^{1/4}},
\end{equation}
where $F$ is the total flux of the galaxy and
$\theta = | \vect{\theta} |$ (where $\vect{\theta}$
is the position on the sky relative to the centre of
the galaxy).
Here $\theta_{\rmn g} = R_{\rmn g} / d_{\rmn A} (0, z_{\rmn g})$,
where $d_{\rmn A}$ is the angular diameter distance to the
galaxy and $R_{\rmn g}$ its intrinsic half-light or effective radius.
This scales with the galaxy's velocity dispersion as
\begin{equation}
\label{equation:g scaling_2df}
R_{\rmn g} = R_{\rmn g*} \left(\frac{\sigma}{\sigma_*}\right)^{u_{\rmn g}},
\end{equation}
where $R_{\rmn g*} = (3 \pm 1)$ kpc and $u_{\rmn g} = 4 \pm 1$
for Es and S0s (Kormendy \& Djorgovski 1989) and
$R_{\rmn g*} = (4 \pm 1)$ kpc and $u_{\rmn g} = 3 \pm 1$
for spirals (de Vaucouleurs \& Pence 1978).

The observed surface brightness profile of a galaxy,
$f_{\rmn g} (\vect{\theta})$, can be found
by convolving the intrinsic profile with the \psf.
Even if the \psf\ is also rotationally-symmetric,
\begin{equation}
\label{equation:sb_prof_2df}
f_{\rmn g} (\theta) =
\end{equation}
\[
\mbox{}
\int_0^\infty \!\! \int_0^{2 \pi}
f_{\rm S\,or\,E} (\theta^\prime)
f_{\rmn see} \left(\sqrt{\theta^2 + \theta^{\prime 2}
- 2 \theta \theta^\prime \cos \theta_\phi^\prime} \, \right)
\theta^\prime {\rmn d} \theta_{\phi}^\prime \,{\rmn d} \theta^\prime,
\]
involves a two-dimensional integral, which must
be evaluated numerically.
This is time-consuming, and so an approximation
was sought.

The analytic surface brightness profile used is 
\begin{eqnarray}
\label{equation:galaxy surface brightness}
f_{\rmn{g}}(\theta)
& \!\!\!\! = \!\!\!\! & 
f_{\rmn g}(0) e^{(\theta^\prime_{\rmn c} / \theta_{\rmn g}^\prime)^\xi}
e^{- \left(\sqrt{\theta^2 + \theta^{\prime 2}_{\rmn c}}
/ \theta_{\rmn g}^\prime \right)^\xi} \\
& \!\!\!\! = \!\!\!\! & 
\frac{\xi F}{2 \pi \theta_{\rmn g}^{\prime 2}
\Gamma(2 / \xi)} \,
\frac{1}{1 - P\left[2/\xi,
(\theta^\prime_{\rmn c} / \theta_{\rmn g}^\prime)^\xi \right]}
e^{- \left(\sqrt{\theta^2 + \theta^{\prime 2}_{\rmn c}}
/ \theta_{\rmn g}^\prime \right)^\xi} , \nonumber
\end{eqnarray}
where $\Gamma(z)$ is the Gamma function
and $P(z, x)$ is the incomplete Gamma function.
Unfortunately $\theta_{\rmn g}^\prime$ can only be found by
solving\begin{equation}
0  =
2 P\left[ 2 / \xi, \left( \sqrt{\theta_{\rmn g}^{\prime 2}
+ \theta_{\rmn c}^2} / \theta_{\rmn g} \right)^\xi \right]
- P\left[2 / \xi, (\theta^\prime_{\rmn c} / \theta_{\rmn g})^\xi \right]
- 1
\end{equation}
numerically, but, more importantly, $f_{\rmn{g}}(\theta)$ can be inverted,
giving
\begin{equation}
\label{equation:th_f}
\theta (f) =
\sqrt{\theta_{\rmn g}^{\prime 2}
\left[ \left(\theta^\prime_{\rmn c}/\theta_{\rmn g}^\prime\right)^\xi
+ \ln \left(f_{\rmn g}(0)/f\right)
\right]^{2/\xi}
- \theta^{\prime 2}_{\rmn c}},
\end{equation}
provided $f \leq f_{\rmn g}(0)$.
The profile can also be integrated, to give the flux within a
specified angle as
\[
F_{\rmn g} (< \theta) =
F
\frac{P\left[2/\xi,
\left(\sqrt{\theta^2 + \theta^{\prime 2}_{\rmn c}}
/ \theta_{\rmn g}^\prime\right)^\xi \right]
- P\left[2/\xi,
(\theta^\prime_{\rmn c} / \theta_{\rmn g}^\prime)^\xi \right]}
{1 - P\left[2/\xi,
(\theta^\prime_{\rmn c} / 2 \theta_{\rmn g}^\prime)^\xi \right]} .
\]
\begin{equation}
\label{equation:f_inside}
\mbox{}
\end{equation}

\begin{figure*}
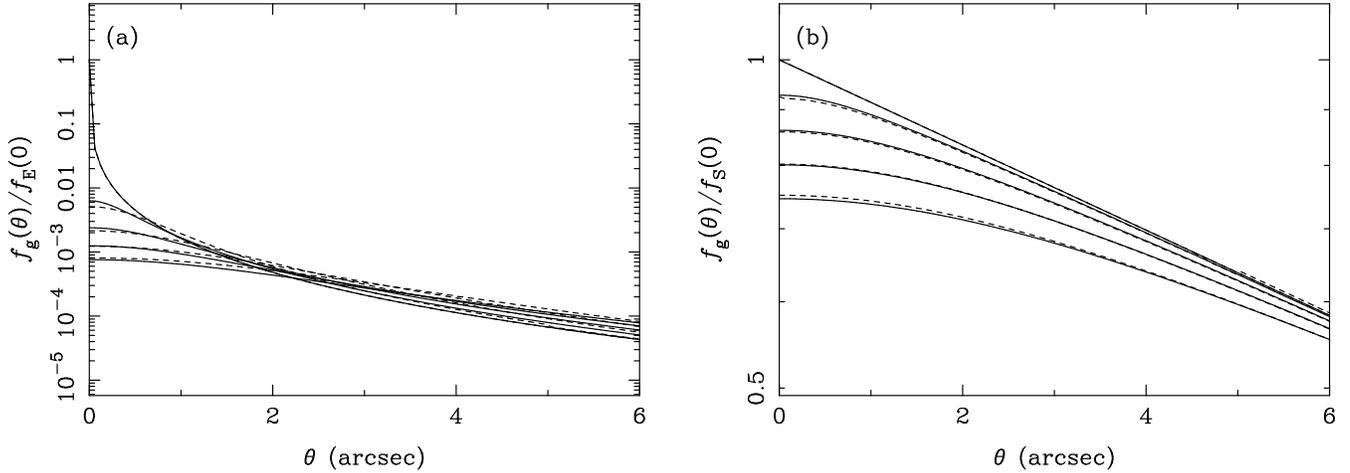

\includegraphics{prof_ell.ps}
\includegraphics{prof_spi.ps}
\vspace{\singlefigureheight}
\caption{The observed surface brightness profiles of
elliptical (a) and spiral (b) galaxies. The dashed lines are
the result of convolving de Vaucouleurs (1948) and Freeman (1970)
profiles, respectively, with a Moffat (1969) \psf;
the solid lines are the analytical approximations
given in \eq{galaxy surface brightness}.
In both cases the galaxy has absolute magnitude $M_* = - 19.4$ and is at a
redshift of 0.1,
and all the profiles are normalised to the peak surface brightness of
the intrinsic galaxy profile. From top to bottom the seeing is
0 arcsec, 1 arcsec, 2 arcsec, 3 arcsec and 4 arcsec.}
\label{figure:profiles}
\end{figure*}

Each of the three parameters plays a
clear role in the shape of $f_{\rmn g} (\theta)$.
The core width, $\theta^\prime_{\rmn c}$, is determined by 
the mainly by the seeing, but has some dependence on the form
of the \psf.
The overall angular scale of the profile, given by
$\theta^\prime_{\rmn g}$, is purely a function of
the galaxy's distance and half-light radius.
For $\theta \gg \theta^\prime_{\rmn c}$ the
surface brightness falls off as
$\exp[(\theta / \theta_{\rmn g})^{- \xi}]$,
assuming that the faint wings of the galaxy are
broader than those of the \psf\footnote{This
is not formally the case for a Moffat (1969) \psf,
although it only becomes relevant on very large
angular scales.}.
Hence $\xi = 1$ for spirals and $\xi = 1/4$ for ellipticals,
and, if $\theta_{\rmn c}^\prime = 0$,
\eq{galaxy surface brightness}
reduces to $f_{\rmn E}$ and $f_{\rmn S}$.
For small $\theta$
\eq{galaxy surface brightness}
tends to $f_{\rmn g} \propto \exp [-
(\theta^2 / 2 \theta_{\rmn c} \theta_{\rmn g}^\prime)^\xi]$, and
so the profile has a Gaussian core only if $\xi \simeq 1$ (i.e.,
for spirals)\footnote{An alternative profile with
\begin{equation}
f_{\rmn g} (\theta) = f_{\rmn g} (0)
e^{\left(\theta_{\rmn c}^\prime / \theta^\prime_{\rmn g}\right)^\xi}
e^{- \left[\left(\theta^{2 / \xi}
+ \theta_{\rmn c}^{\prime 2 / \xi}\right)^{\xi / 2}
/ \theta_{\rmn g}^\prime \right]^\xi}
\end{equation}
would have a Gaussian core for all $\xi$ with the same properties 
as \eq{galaxy surface brightness} for large $\theta$,
but cannot be integrated analytically.}.
\Fig{profiles} shows several examples of
observed galaxies' profiles, comparing the
numerically integrated
curves (dashed lines) to the relevant form of
\eq{galaxy surface brightness} for
both ellipticals (a) and spirals (b).
The agreement for the latter type is excellent,
with relative errors of no more than 1 per cent for the profiles shown.
The average error for the elliptical fits is $\sim 1$ per cent,
and peaks at $\sim 5$ per cent for most of the profiles shown.

\begin{figure*}
\includegraphics{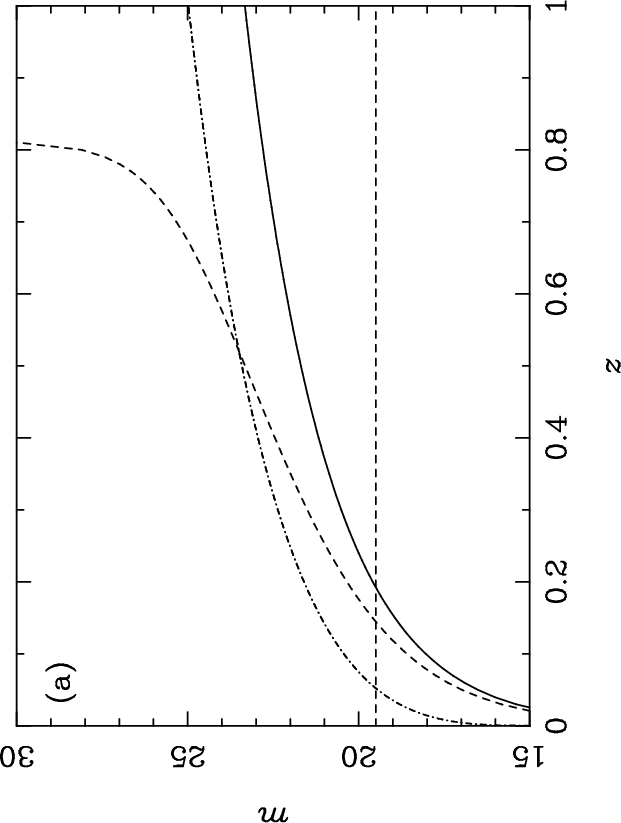}
\includegraphics{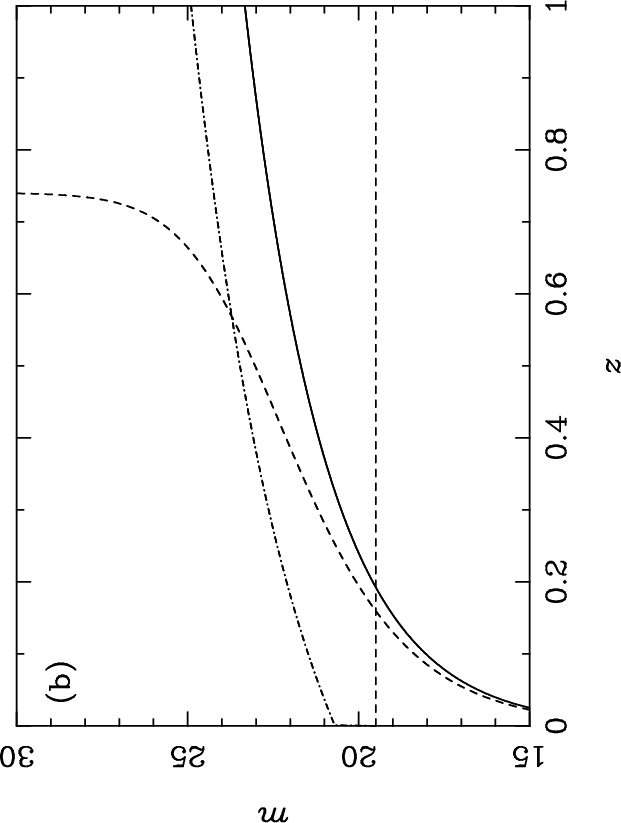}
\vspace{\singlefigureheight}
\caption{The total magnitude, $m$ (solid line),
\APM\ isophotal magnitude, $m^{\prime}$ (dashed line),
and the
fibre magnitude, $m_{\rmn{f}}$ (dot-dashed line),
of an $M_*$ elliptical galaxy (a)
and an $M_*$ spiral galaxy (b),
shown as a function of redshift, $z$.
It is assumed that the
passive luminosity evolution evolution of 
the galaxies cancels out their $K$-corrections.
The horizontal dashed line is the isophotal magnitude limit of the
\TdF\ survey, $m^\prime_{\rm lim} = 19.5$.}
\label{figure:magnitudes}
\end{figure*}

Both the approximate depth of the redshift survey and the 
importance of the use of isophotal magnitudes can be 
gauged by plotting the magnitude of a given galaxy as a function
of redshift (\ie\ a Hubble diagram).
This is shown in \fig{magnitudes}
for both an $M_* = - 19.4$ elliptical and an $M_* = - 19.4$ spiral,
with the \TdF\ fibre magnitude (defined in 
\sect{m_fib_g}) also included.
The $m(z)$ curve is the same for the elliptical and 
the spiral as the luminosity evolution of each type 
is assumed to exactly cancel their $K$-corrections (See Mortlock \& Webster 
2000b), 
but the redshift dependence of the isophotal magnitude
is somewhat different. 
For moderate redshifts, the isophotal flux of 
spirals is actually greater than that of ellipticals,
despite the fact that the latter have more peaked 
surface brightness distributions. 
This comes about as the average flux of 
the spirals over the region registered by 
the \APM\ is higher.
However at high redshifts $m^\prime \rightarrow \infty$ for
all galaxies as they are so faint that their peak surface brightness
is lower than the \APM\ limit of 24.5 mag arcsec$^{-2}$. 
This occurs at a higher redshift for ellipticals,
as it is determined mainly by the central peak
of the surface brightness.
Over the redshift range of the \TdF\ survey (See \sect{dpdz_2df}.)
$m^\prime - m \simeq 0.5$ for most galaxies, and 
so the redshift coverage of the survey is
comparable to that of a \grs\ with
a total limiting magnitude of $m_{\rm lim} \simeq 19$.

\subsection{APM classifications}
\label{section:class_2df}

The observed surface brightness profile 
of a galaxy determines 
its classification as either a stellar or 
non-stellar image. Predictably, almost all the brighter 
galaxies have $\psi_{\rm APM} \ga $ 1000,
and are classified as non-stellar, as shown in \fig{psi_apm}.
It is only fainter than the \TdF\ survey limit of 
$m^\prime = 19.5$ that the galactic and stellar loci begin 
to merge.
The galaxies also occupy a well-defined locus on the
$k_{\rm APM}$-$\mu_{\rm APM}$ plot, as shown in \fig{k-mu}.
Isolated galaxies are monolithic and hence have 
$k_{\rm APM} \simeq 0$, but the cover a wider range in 
$\mu_{\rm APM}$.
Fainter and more distant galaxies become progressively more 
stellar, and so $\mu_{\rm APM} \rightarrow 1$, the 
canonical stellar value. 
Galaxies with $\mu_{\rm APM} \geq 0.8$ are 
in fact classified as stellar mergers,
and rejected from the galaxy sample.
Overall, however, it is clear from \figs{psi_apm} and \ref{figure:k-mu}
that very few real galaxies with $m^\prime \leq 19.5$ are 
lost from the \TdF\ catalogue, and a Monte Carlo simulation 
suggests the \APM\ galaxy survey
has a completeness of $C \sim 95$ per cent, which is comparable
to the values estimated by Maddox \etal\ (1990a).

\subsection{Fibre magnitudes}
\label{section:m_fib_g}

In general, the flux from a survey galaxy that actually enters the 
\TdF\ the spectrograph is considerably
less than the isophotal (or total) flux, 
as the \TdF\ instrument's optical fibres are 
only 1 arcsec in effective radius.
The fibre magnitude of a galaxy of 
total magnitude $m$ is given by
\begin{equation}
m_{\rmn f} = m + 2.5 \log
\left[
\frac{F_{\rmn g}(< \theta_{\rmn f})}{F_{\rmn g}(< \infty)}
\right],
\end{equation}
where $F_{\rmn g}(< \theta)$ is given in \eq{f_inside}.
The $z$-dependence of $m_{\rmn f}$ is shown in 
in \fig{magnitudes} for both an $M_*$ elliptical and 
an $M_*$ spiral galaxy.
The higher central surface brightness 
of the elliptical results in a much higher fraction of its flux 
entering the \TdF\ fibre (dot-dashed line), but the \psf\ 
dominates the observed surface brightness of both types 
by $z \simeq 0.5$, and so there is little difference in $m_{\rmn f}$. 
On average $m_{\rmn f} - m^\prime \simeq 2$ over most 
of the \TdF\ survey's redshift range (\sect{dpdz_2df}),
which increases the survey's potential sensitivity to 
lensed quasars by $\sim 2$ mag. 
However this is somewhat offset by an analogous reduction in the
fibre flux of any lensed quasars behind the survey galaxies; 
this is discussed further in \sect{m_fib}.

\subsection{Redshift distribution}
\label{section:dpdz_2df}

\begin{figure}
\includegraphics{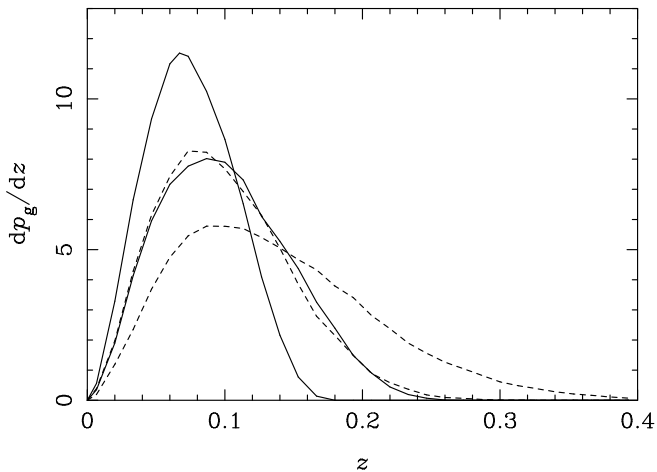}
\vspace{\singlefigureheight}
\caption{The expected redshift distribution, 
${\rmn d}p_{\rmn g}/{\rmn d}z$, of
the galaxies in the \TdF\ \grs. 
The solid lines show the distributions generated if
the correct isophotal magnitude limit of 
$m^\prime_{\rmn lim} = 19.5$ is used;
the dashed lines show the distributions 
implied by a total magnitude limit of 
$m_{\rmn lim} = 19.5$.
In both cases the broader distribution
assumes that the $K$-corrections are exactly cancelled 
by passive luminosity evolution of the galaxies;
the narrower distribution assumes no luminosity evolution,
and uses the $K$-corrections measured by
Kinney \etal\ (1996).}
\label{figure:dpdz_2df}
\end{figure}

The redshift distribution of the galaxies in the \TdF\ \grs\ 
is determined only by the (assumed) intrinsic galaxy population
and the isophotal magnitude limit. In particular,
is independent of the morphological selection criteria of the 
\APM\ survey.
The redshift distribution of a generic, magnitude-limited
galaxy survey was discussed in Mortlock \& Webster (2000b), 
and the results presented here were obtained by the same
method, but with isophotal magnitudes $m^\prime$ replacing 
total magnitudes $m$ throughout. 
Note that this incurs an additional computational overhead,
as the conversion from isophotal magnitude (and redshift)
to absolute magnitude requires an iterative solution.
The resultant redshift distributions, ${\rm d}p_{\rmn g}/{\rm d}z$, 
are shown as solid lines in \fig{dpdz_2df};
the distributions that would be obtained if 
total magnitudes were used are shown as dashed lines.
In both cases the narrower distribution was generated using the 
$K$-corrections of Kinney \etal\ (1996); 
the broader distributions were obtained by
assuming that the $K$-corrections are cancelled by
passive luminosity evolution of galaxies.
The latter provides a 
reasonably simple model of the local galaxy population that
reproduces the observed counts to $m \simeq 23$,
and also matches the observed redshift distribution of the 
\TdF\ galaxies (\eg\ Folkes \etal\ 1999).
The use of isophotal magnitudes in place of 
total magnitudes has almost exactly the same effect as 
omitting the luminosity evolution, relative to the default
model used in Mortlock \& Webster (2000b).
The \TdF\ survey thus has the
same depth as a \grs\ with a total magnitude limit of 
$m \simeq 19$ (as opposed to 19.5).
Using the generic results of 
Mortlock \& Webster (2000b),
this reduction in depth then implies a one third
reduction in the number of lenses in the \TdF\ survey.

\section{Merged objects}
\label{section:merge}

Images separated by 
$\la 5$ arcsec are treated as single objects by the 
\APM\ software, and thus Maddox \etal\ (1990a) introduced 
the $k_{\rmn APM}$-$\mu_{\rmn APM}$ plot to distinguish 
merged images from isolated stars and galaxies. 
The ideas that underpin this method were explained in 
\sect{apm survey}; here the populations of mergers shown 
in \figs{psi_apm} and \ref{figure:k-mu} are defined.

A close pair of images of specified types
are defined by their angular separation,
$\Delta \theta$, and their total magnitudes\footnote{Despite the
fact that isophotal magnitudes are the `natural' choice 
for simulations of the \APM\ survey, the separate isophotal 
magnitudes of two merged objects are somewhat ill-defined.
The total flux of an image pair is simply the sum of the
two components' total fluxes, but their combined isophotal flux 
cannot be determined from their individual isophotal 
fluxes.},
$m_1$ and $m_2$.
For chance alignments, the magnitudes are uncorrelated, 
and so drawn at random from the respective populations, 
and the distribution of angular separation is given
by ${\rmn d}p/{\rmn d}\Delta \theta \propto \Delta \theta$
between 0 and $\Delta \theta_{\rmn max} \simeq 5$ arcsec.
The more complex separation distributions of 
binary stars and physically-associated galaxies are
ignored here, as it is only the region of phase space 
inhabited by merged images that is of interest.

Merged image pairs are shown in 
\figs{psi_apm} and \ref{figure:k-mu} as 
open symbols. Star-star merges are shown as open circles;
star-galaxy mergers as open squares
and galaxy-galaxy mergers as open triangles.
All three populations have $\psi_{\rm APM} \ga $ 1000, 
and lie a significant distant from the stellar 
locus in \fig{psi_apm}. Whilst they also tend to be slightly 
below the galactic population, they would be
treated as extended objects in the \APM\ survey if not
for the use of the $k_{\rm APM}$-$\mu_{\rm APM}$ plot 
(\fig{k-mu}).
Not only are the mergers are quite well separated from the 
monolithic objects (\ie\ single galaxies and stars with
$k_{\rmn APM} \simeq 0$), but the stellar images 
(either isolated stars or stellar mergers) are quite
distinct from those with strong extended components.
This separation allowed Maddox \etal\ (1990a) to
reject most stellar mergers from the galaxy sample, 
whilst retaining most of the galactic mergers.
Unfortunately, many gravitational lenses would also have been rejected,
as discussed below.

\section{Lensed quasars}
\label{section:lens_2df}

The determination of the number of lenses that enter the \TdF\ \grs\ 
splits neatly into two parts: 
a reasonably standard calculation of the lens population;
and the application of the survey selection effects.

The lens calculation uses the same models and methods 
discussed in Mortlock \& Webster (2000b):
the deflectors are assumed to be a non-evolving population
of isothermal spheres 
(with the possibility of both a core radius and ellipticity
accounted for in \sect{results_model}); 
the quasar population is chosen to match the number counts of 
Boyle, Shanks \& Peterson (1988) and the redshift distribution
of the Large Bright Quasar Survey 
(Hewett, Foltz \& Chaffee 1995); and conventional methods are 
used to evaluate the lensing probability (\eg\ Kochanek 1996).

More complex, and more interesting, are the effects of the 
various \APM\ selection effects on the lens statistics.
To enter the \TdF\ \grs, a lensed quasar must satisfy the following
conditions (\cf\ Mortlock \& Webster 2000b):
\begin{enumerate}
\item{The composite object defined by the quasar images and 
	lens galaxy must have an isophotal magnitude $m^\prime <
	m^\prime_{\rmn lim} = 19.5$. This is not a 
	simple function of galaxy and quasar magnitudes,
	but must be computed numerically for each lens.
	Fortunately the quasar images are within the 
	bright (\ie\ $\mu \ll \mu_{\rm lim}$) `core' of 
	the galaxy in many cases, and their total --
	rather than isophotal -- flux can
	be added to the isophotal flux of the galaxy.}
\item{The quasar must be bright enough to for its
	spectroscopic signature to be present in the 
	composite spectrum obtained during the \TdF\ \grs.
	Following Kochanek (1992),
	the quasar is assumed to be detectable if the 
	fibre magnitudes of the quasar images and galaxy
	satisfy $m_{\rmn f,q} - m_{\rmn f,g} \leq 
	\Delta m_{\rmn f,qg} \simeq 2$. 
	The explicit calculation of the fibre
	magnitudes is discussed in \sects{m_fib_g} and
	\ref{section:m_fib}.}
\item{The composite object must be classed as a galaxy 
	according to the \APM\ star-galaxy separation
	scheme. This 
	condition is taken to be satisfied if 
	$m_{\rmn g} - m_{\rmn q} \leq \Delta m_{\rmn gq}$, 
	where the value of $\Delta m_{\rmn gq}$ is estimated
	in \sect{lens_apm_class}. Note that $m_{\rmn q}$ is 
	the total, lensed magnitude of all the quasar images.}
\end{enumerate}
The small aperture size tends to result in a
greater sensitivity to the more centralised quasar images 
(\sect{m_fib}),
but
the use of isophotal magnitudes decreases the effective depth
of the survey, and the rejection of 
multiple point sources also tends to reduce the
number of lenses (\sect{lens_apm_class}).
The relative importance of these competing effects cannot be
judged a priori, but must be determined numerically, 
as discussed in \sect{results_2df}. 

\subsection{Fibre magnitudes}
\label{section:m_fib}

\Fig{magnitudes} shows that only a small fraction of the 
light from most galaxies enters the small \TdF\ 
optical fibres. However, much of the light from a lensed
quasar can also fall outside the fibre, both
due to seeing effects and the fact that the quasar images
are generally somewhat offset from the centre of the fibre.

The flux from a lensed quasar which enters the fibre
is calculated by integrating the surface brightness of 
each quasar image over the fibre, and summing the resultant fibre fluxes. 
Thus the fibre magnitude, $m_{\rmn f,q}$, of a lensed quasar with source
position $\vect{\beta}$ is
\begin{equation}
\label{equation:m_fq}
m_{\rmn{f,q}} = m_{\rmn q} - 2.5 \log \left\{
\frac{
\sum_i \mu \left[\vect{\theta}_i(\vect{\beta}) \right] \,
\Upsilon[\vect{\theta}_i(\vect{\beta})]
}
{
\sum_i \mu \left[\vect{\theta}_i(\vect{\beta}) \right] 
}
\right\} ,
\end{equation}
where $\Upsilon (\vect{\theta})$ is the fraction of the light
from an image at $\vect{\theta}$ which enters the fibre.
The image positions, $\vect{\theta}_i$, and magnifications,
$\mu_i$, are functions only of source position, $\vect{\beta}$,
and are regarded as inputs to this calculation.
Performing the two-dimensional integrals over 
the aperture can be time-consuming, 
but, for simple \psfs\ [including \eq{moffat}],
the problem can be 
reduced to a one-dimensional integral (Mortlock 1999).

\begin{figure*}
\includegraphics{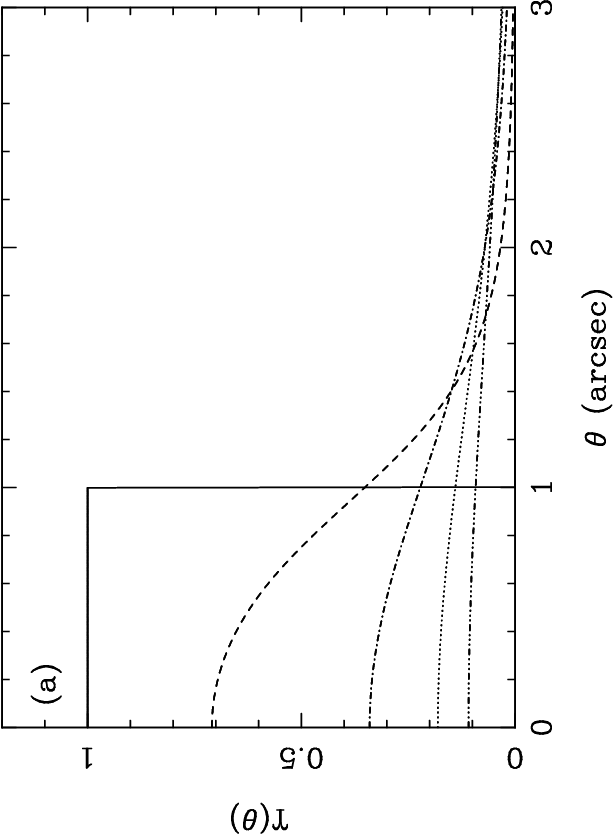}
\includegraphics{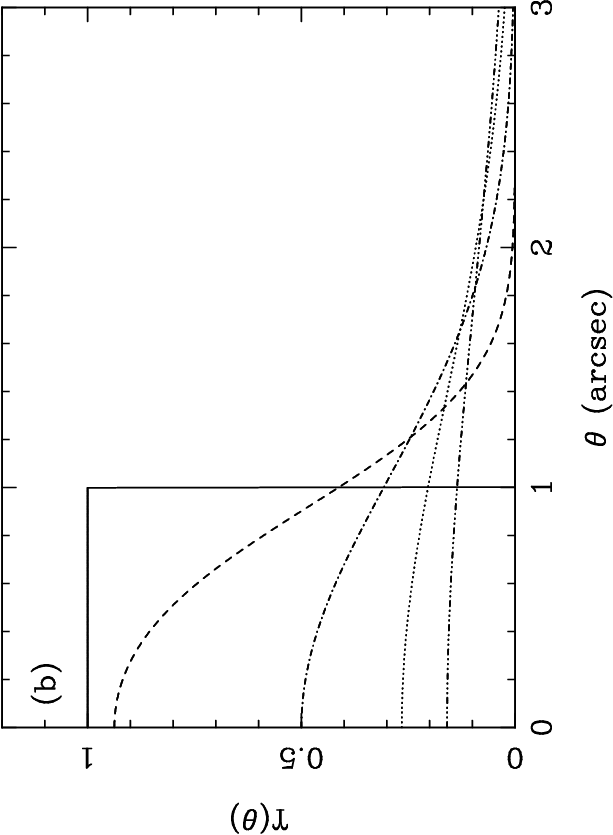}
\vspace{\singlefigureheight}
\caption{The fraction of the light, $\Upsilon$, from a point-like image
that enters a \TdF\
optical fibre as a function of $\theta$, the
angular separation between the centre of the image and the centre of the
1 arcsec radius fibre.
A Moffat (1969) \psf\ is assumed, with $\eta = 2.5$ in (a) and
$\eta \rightarrow \infty$ (Gaussian seeing) in (b).
In both panels $\Upsilon$ is shown for a several values of the
seeing:
$\theta_{\rmn s} = 0$ arcsec (solid line);
$\theta_{\rmn s} = 1$ arcsec (dashed line);
$\theta_{\rmn s} = 2$ arcsec (dot-dashed line);
$\theta_{\rmn s} = 3$ arcsec (dotted line);
and
$\theta_{\rmn s} = 4$ arcsec (dot-dot-dot-dashed line).}
\label{figure:fibre}
\end{figure*}

The dependence of $\Upsilon$ on both $\theta_i$ and 
$\theta_{\rmn{s}}$ is shown in \fig{fibre}.
The solid line in each plot
shows the step function obtained if the \psf\ is ignored.
The other four curves show that an appreciable portion
of the flux from
images up to $\sim 2$ arcsec from the centre of the fibre is
registered in the worst seeing to be expected at the \AAT.
This effect is more important for \psfs\ with more
pronounced wings,
as can be seen by comparing the results for
the `default' Moffat (1969) profile ($\eta = 2.5$) shown in 
\fig{fibre} (a) with those for Gaussian seeing 
($\eta \rightarrow \infty$) shown in (b).

\begin{figure*}
\includegraphics{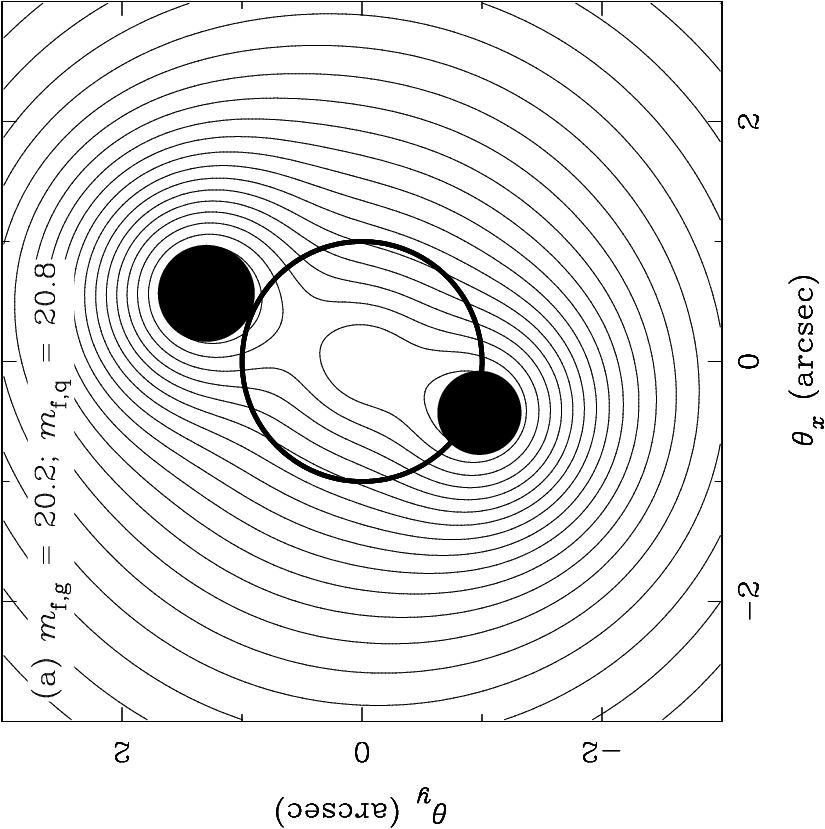}
\includegraphics{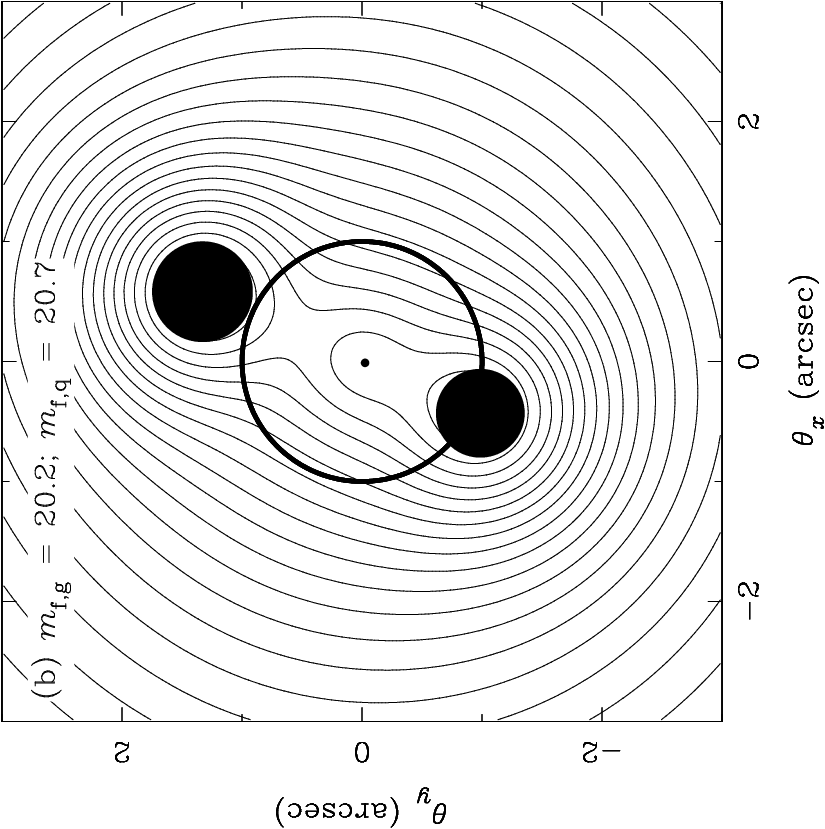}
\includegraphics{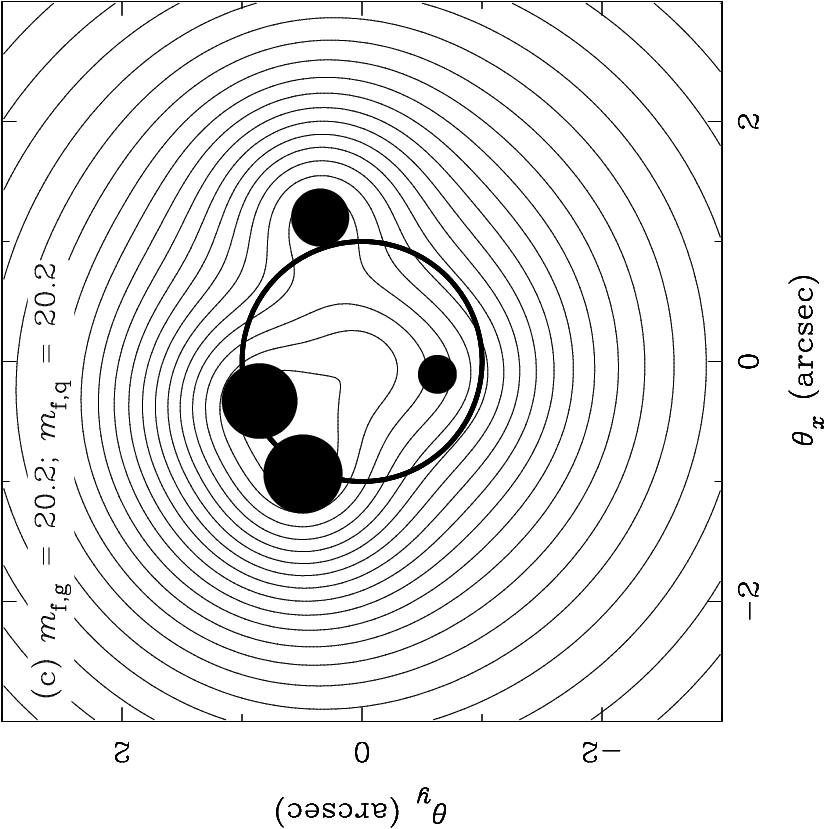}
\includegraphics{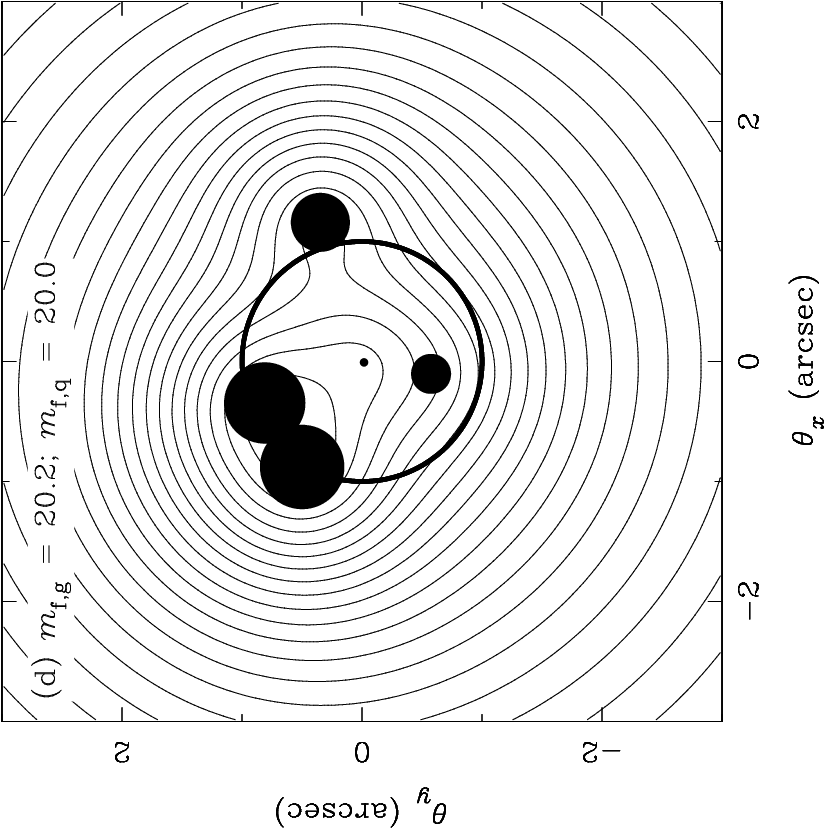}
\vspace{18.2cm}
\caption{Simulated images of a $\sigma_* = 220$ km s$^{-1}$ elliptical galaxy
(at $z_{\rmn g} = 0.1$ and with $m_{\rmn g} = 17.9$)
lensing a $m_{\rmn q} = 22$ quasar at $z_{\rmn q} = 2$.
The galaxy has $m^\prime \simeq 18.5$ in each of the panels,
which are labelled
with the fibre magnitudes of the galaxy, $m_{\rmn f,g}$, and
the quasar, $m_{\rmn f,q}$.
In all cases the quasar is off axis by $\sim 0.2$ arcsec, and 
its images are shown by the black disks, the areas of which
are proportional to their magnifications. The heavy circle is the
\TdF\ fibre, and the contours are uniformly spaced in 
logarithmic surface brightness.
In (a) the galaxy is a singular isothermal sphere;
in (b) it has a core radius of $r_{\rmn c} = 100$ pc;
in (c) there is an external shear of $\gamma_0 = 0.2$ 
(equivalent to an ellipticity of $e \simeq 0.6$);
and in (d) the galaxy has both a core radius of $r_{\rmn c} = 100$ pc
and an external shear of $\gamma_0 = 0.2$.
In all cases the seeing is $\theta_{\rmn s} = 1$ arcsec,
and it is assumed that the night sky has zero surface brightness.}
\label{figure:sky}
\end{figure*}

Irrespective of the details of the \psf,
\fig{fibre} illustrates how the positions of the quasar images
can influence $m_{\rmn f,q}$,
and thus the chance of a lens being detected in the \grs. 
Obviously the $\vect{\theta}_i$s are determined by $\vect{\beta}$,
but the types of image configurations produced depend on the 
mass distrbution of the deflector.
Several representative examples of this are shown in \fig{sky},
in which the surface brightness of a galaxy-quasar composite 
is shown for a number of plausible lens models. 
The galaxy is an $M_*$ elliptical at a redshift of 0.1,
and has $m_{\rmn g} = 17.9$ and $m^\prime_{\rmn g} = 18.5$, ignoring the 
quasar light.
In all panels the quasar is slightly mis-aligned with the galaxy,
and has an unlensed magnitude of 22. 
The assumption of 1 arcsec seeing means that the 
quasar images are sufficiently smeared
that they have only a gross influence on the surface
brightness contours, and often remain unresolved.

The default mass model, a singular isothermal sphere (as used for most of 
the calculations discussed in \sect{results_2df}),
is shown in (a). The two images lie just outside the fibre,
but the \psf\ is sufficiently broad that the fibre flux of the
quasar is three times its unlensed flux.

In (b) the galaxy is no longer singular -- the density flattens off
within $r_{\rmn c} \simeq 100$ pc of its centre, resulting in the formation of 
an extra image and generally higher magnifications.
This results in an increased fibre flux, but, for reasonable core radii,
the effect is only minimal -- just 10 per cent in the example shown.
Note that there is no significant change in the image 
separation for small $r_{\rmn c}$ 
-- use of a self-consistent dynamical normalisation 
(Kochanek 1996; Mortlock \& Webster 2000a) cancels out the 
reduced depth of the galaxy's central potential well.
For very large cores, 
the image separation actually increases, and hardly any of the 
quasar's light would enter the fibre.
As shown in \fig{n_lens_2df_lens} (a) this reduces the lensing
probability to zero.
The probability of lensing by a spiral galaxy has a similar dependence on 
the core radius, 
but for very different reasons.
The dynamical normalisation of spirals is taken from their rotation
curves, which are essentially unaffected by the core structure.
Hence increasing $r_{\rmn c}$ results 
in reduced image separations, and an even greater increase in the quasar's 
fibre flux. However the shrinking of the radial caustic 
dominates the statistics, again reducing the lensing probability
to zero for large cores.

More important than the core structure of the deflector is its 
ellipticity. A mass profile of ellipticity of $e$ can be modelled
adequately by applying an external shear of magnitude $\gamma_0 = e / 3$
(\eg\ Kassiola \& Kovner 1993), 
and \fig{sky} (c) shows a singular lens with $\gamma_0 = 0.2$,
representing a galaxy of slightly greater than average ellipticity
(\eg\ Keeton, Kochanek \& Seljak 1997). 
The source position was chosen so that it lies
inside the lens's tangential caustic, and the 
extra pair of images contribute significantly to the detectability
of the lens.
The total magnification of the source is $\sim 16$, and the 
quasar's fibre magnitude is $20.2$.
Extra magnification is a generic feature of quadrupole lenses
(\eg\ Kochanek \& Blandford 1987; Keeton \etal\ 1997), 
and the inclusion of ellipticity in the calculation
can increase the expected number of lenses by up to 50 per cent
[\fig{n_lens_2df_lens} (b)]. 

Finally, \fig{sky} (d) shows the combined effect of a core radius
and finite ellipticity -- relative to the singular model, the 
magnifications are slightly increased and there is an extra image,
which increases the quasar's fibre flux by a further 20 per cent. 
It can also been seen that, in rough terms, the effects of 
core radius and deflector ellipticity decouple, and it is the
shape of the lens that dominates the lens statistics.

Note that in all four of the above examples 
the overall magnification is high enough to
ensure the lensed quasar's fibre flux is greater
than its total unlensed flux would have been. 
Nonetheless, for most lenses
more than half the light of the lensed quasar images falls outside 
the fibre -- typically
$m_{\rmn f,q} - m_{\rmn q} \simeq 1$.
This is a significant loss, but
the effect is still much smaller than it is for galaxies,
which have $m_{\rmn f,g} - m_{\rmn g} \simeq 2$ on average (\sect{m_fib_g}).
Thus the small diameter of the \TdF\ fibres is 
favourable for spectroscopic lens searches.

\subsection{APM classifications}
\label{section:lens_apm_class}

The \APM\ classification of a lensed quasar
as stellar, merged or non-stellar is a function
of its surface brightness, and hence sensitive to both
the properties of the lens galaxy and the positions and
magnifications of the quasar images.
The number of independent variables that `define' a
given lens is too high to facilitate an analytic
investigation, and so 
Monte Carlo simulations of
a large number of source-deflector pairs were generated.
%The lens model used was the singular isothermal sphere
%with an external shear of $\gamma_0 = 0.1$.
For each lens, the three \APM\ image classification
parameters ($\psi_{\rm APM}$, $k_{\rm APM}$ and $\mu_{\rm APM}$,
as described in \sect{apm survey}) were calculated,
and the simulated lenses are plotted as filled circles
in \figs{psi_apm} and \ref{figure:k-mu}.

As expected, almost all the lenses have $\psi_{\rm APM} \ga $ 1000,
indicating that their surface brightness profiles are significantly
different from those of stars.
However the population of lenses is quite similar to the
merger population discussed in \sect{merge}. In particular,
two-image lenses with high-redshift deflectors 
are almost indistinguishable from binary stars.
Hence the rejection of star-star mergers from the \APM\
galaxy survey on the basis of their location in
$k_{\rm APM}$-$\mu_{\rm APM}$ space also results in many
lenses being lost from the \TdF\ \grs.
In fact the only lenses that remain in significant numbers
are those with bright (and hence nearby) lens galaxies.
Almost all lenses with $m_{\rmn g} - m_{\rmn q} \la -1$
are morphologically classified as galaxies, and 
those with $m_{\rmn g} - m_{\rmn q} \ga 1$ are flagged as 
stellar mergers. 
In the calculation that 
follows, this is simplified to the requirement that all lenses satisfy
$m_{\rmn g} - m_{\rmn q} \leq \Delta m_{\rmn gq} = 0 \pm 1$.

%\Fig{delta_m_2df} shows how the classification of the lenses
%varies with the total magnitude of the source-deflector
%composite, $m$, and the magnitude difference between the
%two components, $m_{\rmn g} - m_{\rmn q}$.
%Whilst there is no clear break, this simulation suggests that
%$\Delta m_{\rmn gq} \simeq 2$ for the \TdF\ galaxy survey.
%
%\begin{figure}
%\special{psfile=delta_m_2df.ps angle=-90 vscale=60 hscale=60
%	hoffset=-15 voffset=28}
%\vspace{\singlefigureheight}
%\caption{A number of simulated lensed quasars as a function of the
%total magnitude of the source-deflector
%composite, $m$, and the magnitude difference between the
%two components, $m_{\rmn g} - m_{\rmn q}$. The different
%symbols denote the \APM\ classifications:
%stellar (points); merged images (open circles);
%and non-stellar (\ie\ galactic; filled circles).}
%\label{figure:delta_m_2df}
%\end{figure}

\section{Results}
\label{section:results_2df}

The aim of this calculation is essentially to find 
one value: 
$N_{\rmn l}$, the number of lensed quasars expected in the \TdF\ \grs.
This is given by integrating the lensing probability of 
individual galaxies over the entire galaxy population, subject 
to the selection effects described in \sect{lens_2df}.
In \sect{results_sel} the dependence of 
$N_{\rmn l}$ on the survey design is investigated;
the variation with lens model is discussed in \sect{results_model}.

\subsection{Survey design}
\label{section:results_sel}

\begin{figure*}
\includegraphics{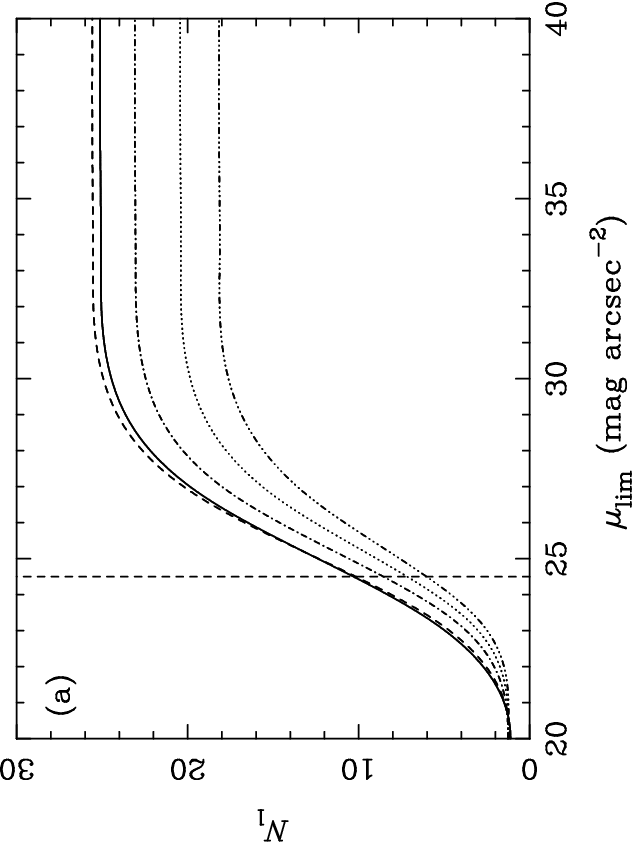}
\includegraphics{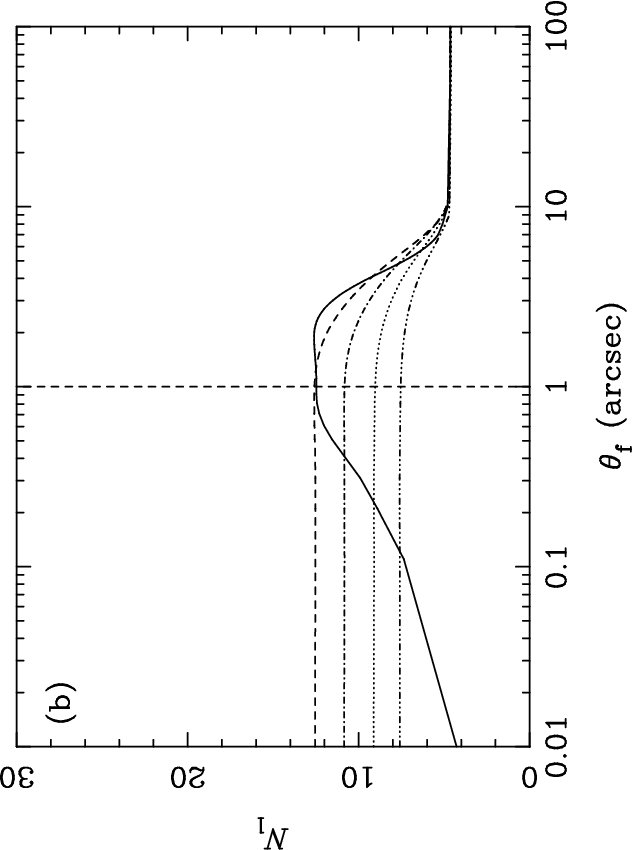}
\includegraphics{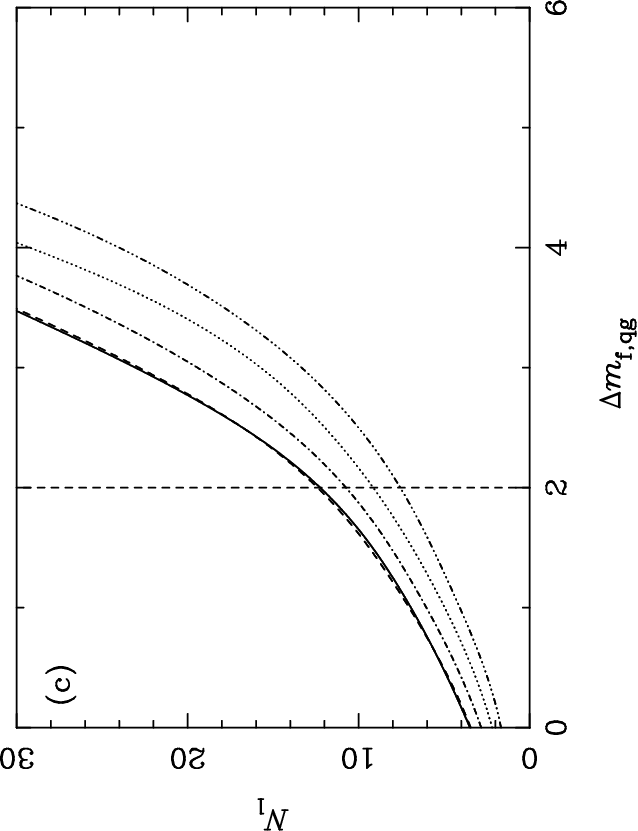}
\includegraphics{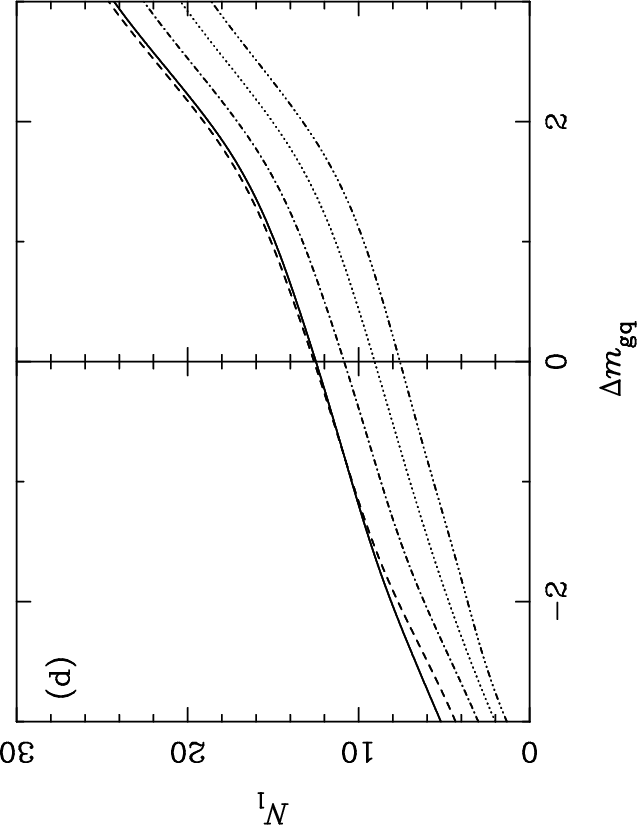}
\vspace{\doublefigureheight}
\caption{The expected number of lenses, $N_{\rmn l}$, in a \grs\ of the same 
design as the \TdF\ survey, as a function of
limiting surface brightness, $\mu_{\rmn lim}$, in (a),
fibre radius, $\theta_{\rmn f}$, in (b), 
$\Delta m_{\rmn f,qg}$ in (c)
and $\Delta m_{\rmn gq}$ in (d).
The vertical dashed lines indicate the values appropriate to
the \TdF\ \grs: $\mu_{\rmn lim} = 24.5$, $\theta_{\rmn f} = 
1$ arcsec, $\Delta m_{\rmn f,qg} = 2$ and $\Delta m_{\rmn f,qg} = 0$.
The lens model used was the singular isothermal sphere with no 
external shear.
The results are shown for several values of the atmospheric 
seeing:
$\theta_{\rmn s} = 0$ arcsec (solid lines);
$\theta_{\rmn s} = 1$ arcsec (dashed lines);
$\theta_{\rmn s} = 2$ arcsec (dot-dashed lines);
$\theta_{\rmn s} = 3$ arcsec (dotted lines);
and
$\theta_{\rmn s} = 4$ arcsec (dot-dot-dot-dashed lines).}
\label{figure:n_lens_2df_obs}
\end{figure*}

The generic calculation of lensing in redshift surveys
presented in Mortlock \& Webster (2000b) implies that
$\sim 10$ lenses would be discovered in a survey of
$N_{\rmn tot} = 2.5 \times 10^5$ galaxies that is complete
to a total magnitude limit of $m = 19.5$.
Implicit in this calculation were the assumptions that
there are no surface brightness-related selection
effects in the survey, and also
that the apertures used to gather the survey spectra are
larger than the typical angular scales of both the galaxies and lensed
images.
These assumptions are dropped here, 
and \fig{n_lens_2df_obs} shows how
$N_{\rmn l}$ varies with various aspects of the survey design;
the generic model can be reproduced by setting $\mu_{\rmn lim}
\rightarrow \infty$ and $\theta_{\rmn f} \rightarrow \infty$.
In this limiting case, both
the fibre magnitude, $m_{\rmn f}$, 
and its isophotal magnitude, $m^\prime$, of an object
approach its total magnitude, $m$.

\Fig{n_lens_2df_obs} (a) shows how $N_{\rmn l}$ increases with
$\mu_{\rmn lim}$, the isophotal limit of the survey. 
Most galaxies 
with $m \simeq 19.5$ 
have observed peak surface brightnesses
close to $22$ mag arcsec$^{-2}$, and so the properties of 
the redshift survey (and $N_{\rmn l}$ in particular) do not 
vary with $\mu_{\rmn lim}$ if the isophotal limit is 
either much smaller than or much greater than this critical value.
The increase of $N_{\rmn l}$ with $\mu_{\rmn lim}$ is simply
due to the increased mean redshift of the survey galaxy population,
as described in \sect{dpdz_2df}.
However for $\mu_{\rmn lim} \ga 30$ the isophotal and total
magnitudes of all the objects considered here are essentially the same,
and $N_{\rmn l}$ flattens off.

Interestingly, in the limiting case of $\mu_{\rmn lim} \rightarrow \infty$,
$N_{\rmn l}$ is twice the value obtained in Mortlock \& Webster (2000b).
This is due to the finite aperture size, as shown in
\fig{n_lens_2df_obs} (b). 
Taking $\theta_{\rmn f} \rightarrow \infty$ 
halves $N_{\rmn l}$,
as the increase in the fibre flux of the survey galaxies is
much greater than that of the more centralised quasar images.
However, $N_{\rmn l}$, does not increase for 
$\theta_{\rmn f} \la 1$ arcsec, as there are few lenses 
with image separations less than $\sim 2$ arcsec.
This implies that the \TdF\ fibre size is close to optimal for a lens
search: 
any smaller and the integration times required to obtain
reasonable spectra of $m \simeq 19$ galaxies would 
become prohibatively long;
any larger and fewer lenses would be found.

Also of interest is the variation of $N_{\rmn l}$ 
with the seeing, which is illustrated by the five 
lines in each panel of \figs{n_lens_2df_obs} 
and \ref{figure:n_lens_2df_lens}. In most cases 
there is little difference between 
$\theta_{\rmn s} = 0$ arcsec and 
$\theta_{\rmn s} = 1$ arcsec, primarily because 
$\theta_{\rmn s} \la \theta_{\rmn f}$ in this range. 
However if $\theta_{\rmn s} \ga \theta_{\rmn f}$
the number of lenses decreases with the 
seeing, because
so much of the flux from even centred quasar 
images misses the fibres, as illustrated in 
\fig{fibre}. (The survey galaxies are extended intrinsically,
and so the relative reduction in fibre flux is lower.)
The one exception to this qualitative explanation
is shown in \fig{n_lens_2df_obs} (b), in which the
number of lenses decreases rapidly with decreasing 
fibre size when $\theta_{\rmn s} = 0$. 
This case can be understood geometrically, as 
the quasar images must lie inside the \TdF\ fibres to 
be registered at all (there being no `spillage' from the
\psf), whereas most low-redshift galaxies produce image
separations slightly larger than the fibre diameter.

\Fig{n_lens_2df_obs} (c) shows an exponential 
increase in $N_{\rmn l}$ with $\Delta m_{\rmn f,qg}$, which 
reflects the steepness of the quasar luminosity function. 
For $\Delta m_{\rmn f,qg} \ga 2$ the change in 
$\Delta m_{\rmn f,qg}$ corresponds almost directly to the 
increase in the effective depth of the `lens survey'. 
It is also clear that $N_{\rmn l}$ is more sensitive to
$\Delta m_{\rmn f,qg}$ than it is to any other model parameter. 
If a detailed analysis of the \TdF\ spectra revealed 
that $\Delta m_{\rmn f,qg} \simeq 2$ was an underestimate of 
the quasars' `spectral prominence', the expected 
number of lenses could easily be doubled.

As shown in \fig{n_lens_2df_obs} (d),
the $\Delta m_{\rmn gq}$ dependence is similar, but less
pronounced than the dependence on $\Delta m_{\rmn f,qg}$ discussed 
above. This is because a high value of $\Delta m_{\rmn gq}$
only results in the `lens survey' being more complete to 
a given depth, whereas increasing $\Delta m_{\rmn f,qg}$
probes fainter magnitudes and thus larger numbers of quasars.
Combined with the fact that $\Delta m_{\rmn gq}$ is 
reliably constrained to be close to 0 (from \sect{lens_apm_class}), 
the uncertainties 
in the star-galaxy separation properties of the \APM\ survey
do not flow through to the calculation of $N_{\rmn l}$.

\subsection{Lens model}
\label{section:results_model}

The most significant uncertainties in the calculation
of $N_{\rmn l}$ are related to the observational 
parameters discussed in \sect{results_sel}, but
the results do vary with the lens model,
and the effects of a finite core radius and 
ellipticity were explored.

\begin{figure*}
\includegraphics{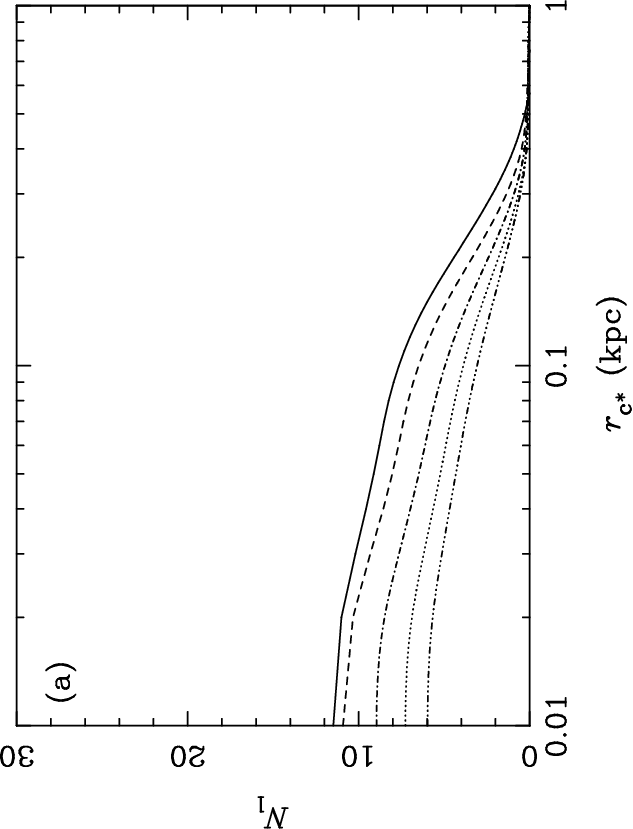}
\includegraphics{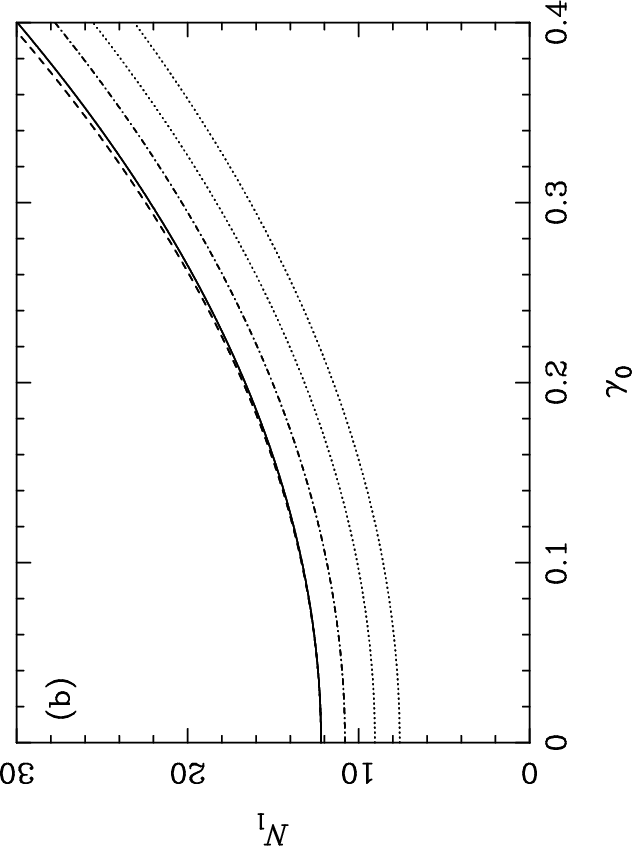}
\vspace{\singlefigureheight}
\caption{The expected number of lenses in the \TdF\ \grs, $N_{\rmn l}$,
as a function of
the the canonical core radius of the lens galaxies, $r_{\rmn c*}$, in (a)
and the external shear, $\gamma_0$, in (b).
The calculation is performed 
assuming
$\Delta m_{\rmn f,qg} = 2$
and
$\Delta m_{\rmn gq} = 0$ in all cases.
The results are shown for several values of the atmospheric
seeing:
$\theta_{\rmn s} = 0$ arcsec (solid lines);
$\theta_{\rmn s} = 1$ arcsec (dashed lines);
$\theta_{\rmn s} = 2$ arcsec (dot-dashed lines);
$\theta_{\rmn s} = 3$ arcsec (dotted lines);
and
$\theta_{\rmn s} = 4$ arcsec (dot-dot-dot-dashed lines).}
\label{figure:n_lens_2df_lens}
\end{figure*}

\Fig{n_lens_2df_lens} (a) shows how 
$N_{\rmn l}$ varies with canonical core radius, 
$r_{\rmn c*}$. The core radius is assumed to vary with 
velocity dispersion as $r_{\rmn c} = r_{\rmn c*} (\sigma / \sigma_*)^4$, 
and $r_{\rmn c*}$ is taken to be the same for each galaxy
type. 
The fall-off of $N_{\rmn l}$ with 
core radius is very pronounced, as explained in \sect{m_fib} --
for large $r_{\rmn c*}$ the spirals' cross-sections become zero
and the images formed by Es and S0s completely miss the 
\TdF\ fibres.
So the assumption of a singular lens model overestimates 
$N_{\rmn l}$ by at most
$30$ per cent, given that
$r_{\rmn c*} < 100$ pc (Kochanek 1996).

More important is the ellipticity of the lens galaxies, 
which are typically inferred to be as high as $e \simeq 0.6$ 
from lensing studies (\eg\ Keeton \etal\ 1997).
This was investigated by the application of an external
shear of magnitude $\gamma_0 = e / 3$, with all galaxies 
assumed to have the same projected shape -- a more realistic
distribution would simply smooth the curves shown in 
\fig{n_lens_2df_lens} (b). 
As discussed in \sect{m_fib}
and shown in \fig{sky} (c), the extra images that 
can be produced by quadrupole lenses almost always lie
inside the \TdF\ fibres, and so the greater magnification
bias can significantly increase $N_{\rmn l}$. Given that
$\gamma_0 \ga 0.3$ is unrealistic for the dark matter-dominated
galaxy halos, the use of a circularly-symmetric lens model
underestimates the number of lenses by at most 50 per cent.

Overall, the uncertainty in the deflectors' mass distributions does not 
greatly affect the estimates of $N_{\rmn l}$ as the two effects 
described above tend to cancel out.
The most
realistic model would have a small ($\la 50$ pc)
core radius and an ellipticity of between 0.3 and 0.6
(\ie\ a shear of between 0.1 and 0.2)
which would result in only slightly more lenses than the 
default singular model. 
The calculation using 
singular isothermal spheres also
provides a robust lower limit of $N_{\rmn l} \simeq 10$.

\section{Conclusions}
\label{section:conc_2df}

Kochanek (1992) and Mortlock \& Webster (2000b)
showed that appreciable numbers of lensed quasars 
could be discovered in galaxy surveys, but a number of potentially
important selection effects and biases were ignored in both
calculations. 
These generic results have been refined here to obtain a 
more realistic estimate of the number of lenses, $N_{\rmn l}$,
expected in the \TdF\ \grs. 

In particular, the importance of 
a finite collection area for the redshift survey spectra
and a finite surface brightness limit
were investigated. 
The \TdF\ optical fibres are unusually small (an angular
radius of only 1 arcsec), which approximately
doubles the number of lenses, as so much of the 
light from the extended galaxies misses the fibres.
Conversely, the use of isophotal rather than total magnitudes 
reduces the effective depth of the survey (by $\sim 0.5$ mag
in the case of \TdF)
which reduces the depth of the `lens survey' by a similar amount.

The \TdF\ input catalogue was determined by the 
\APM\ star-galaxy separation algorithm, and 
so this was applied to a simulated population of quasar lenses,
with the result that most were classified as multiple stellar 
images.
Nonetheless, any lens in which 
the deflector galaxy is at least as bright as the (magnified) quasar images
should enter the \APM\ galaxy survey and hence the \TdF\ \grs.
Importantly, lenses with nearby deflectors
(\eg\ Q~2237+0305)
are especially valuable, as discussed in Mortlock \& Webster (2000b).

Whilst each of the above selection effects can change 
$N_{\rmn l}$ by a factor of several, they tend to cancel 
each other out. 
The generic estimate in Mortlock \& Webster (2000b)
implied that between
ten (if only lenses with bright deflectors were included)
and fifty (if lenses with high-redshift deflectors
were not rejected from the survey)
new quasar lenses 
could be discovered in the \TdF\ \grs.
Clearly it the former figure which is most relevant,
and thus the prior prediction of $\sim 10$ low-redshift lenses is 
confirmed by the more detailed simulations presented here.
The random uncertainty in this value is mainly due to shot noise, but
the variability of the atmospheric
seeing at the \AAT\ site is also an issue.
The most important systematic error is the 
quality of the survey spectra (characterised by
$\Delta m_{\rmn f,qg}$), as it is not clear how efficiently
galaxy spectra can be searched for the presence of quasar
emission features.

Irrespective of the above uncertainties, it is clear that
redshift surveys, and the \TdF\ \grs\ in particular,
are potential sources of large numbers of lensed quasars.
The next step then is to implement a systematic search for 
quasars' emission lines amongst the \TdF\ galaxy spectra. 
A notable success in this field
was the spectroscopic discovery by Warren \etal\ (1996) 
of a lensed emission line galaxy. In a related project, 
Willis \etal\ (2000) have developed a more rigorous technique,
using template-subtraction to search for 
lensed emission line galaxies in a \TdF\ sample of field ellipticals.

Unfortunately these methods are probably not applicable to the 
\TdF\ spectra as the target galaxies are not as homogeneous as
those in the above sample.
The most obvious choice of search technique
is principal components
analysis (\eg\ Murtagh \& Hecht 1987), as it is already being used
to analyse the \TdF\ galaxy
spectra (Glazebrook, Offer \& Deeley \etal\ 1999; Folkes \etal\ 1999). 
Further, an important part of the existing analysis software is the 
flagging of `unusual' spectra\footnote{At the time of writing, 
about $10^5$ survey spectra had been obtained, reduced, and processed.
No lenses or lens candidates have been reported, but little effort
has been put into the analysis of irregular objects thus far.
It is not clear whether the
pipeline reduction algorithms will be sufficient to generate a 
useful list of lens candidates; this should become more apparent
as the survey nears completion.},
and it is a subset of these that
will become the lens candidates.
Even if there are thousands of `unusual' objects, it would be feasible
to search for the presence of quasar emission features by eye, 
which would probably yield several hundred genuine lens candidates,
most of which would then be easily rejected on
morphological grounds.
The completeness of the final candidate sample
may well be quite low,
but its efficiency (\ie\ the fraction of lens candidates
which are real) should be very high relative to conventional
lens surveys (Mortlock \& Webster 2000b). 
The discovery of even one lens in this manner would not only
show the worth of such searches, but be valuable in its own
right.

\section*{Acknowledgments}

Matthew Colless and Steve Maddox 
provided a great deal of insight into the 
workings of the \TdF\ redshift survey and the \APM\ survey,
respectively. The comments of the anonymous referee also 
resulted in some significant improvements to this paper.
DJM was supported by an Australian Postgraduate Award.

\bsp
\label{lastpage}
\end{document}